\documentclass[a4paper,11pt
]{article}

\usepackage{amsmath,amssymb,theorem,enumerate}
\usepackage{tikz} 
\usetikzlibrary{positioning, intersections, calc, arrows.meta,math} 

\theorembodyfont{\normalfont\slshape}
\newtheorem{thm}{Theorem}

\newtheorem{Thm}{Theorem}

\newtheorem{cor}[thm]{Corollary}

\newtheorem{lem}[thm]{Lemma}

\newtheorem{obs}[thm]{Observation}

\newtheorem{Q}[Thm]{Question} 
\newtheorem{claim}{Claim}[section]

\newtheorem{con}[thm]{Conjecture}
\newtheorem{Con}[Thm]{Conjecture}

\newtheorem{rem}[thm]{Remark}
\numberwithin{equation}{section}

\newcommand{\qed}{{$\square$\bigbreak}}
\newcommand{\proof}[1]{%
\medbreak\noindent%
{\bfseries Proof{{\bfseries #1}}.\ }%
}

\title{
Contributions to conjectures on planar graphs: Induced Subgraphs, Treewidth, and Dominating Sets
}

\author{
Kengo Enami$^{1}$
\thanks{This work was supported by JSPS KAKENHI Grant Number JP23K13006}
\thanks{E-mail address: \texttt{enamikengo@gmail.com}}
\hspace{+8pt}
Naoki Matsumoto$^{2}$
\thanks{This work was supported by JSPS KAKENHI Grant Number JP22K11911}
\thanks{E-mail address: \texttt{naoki.matsumo10@gmail.com}}
\hspace{+8pt}
Takamasa Yashima$^{3}$
\thanks{This work was partially supported by JSPS KAKENHI Grant Number JP20K14353}
\thanks{E-mail address: \texttt{takamasa.yashima@gmail.com}}
\vspace{+8pt}
 \\
\small
$^1$\small\textsl{
Department of Computer Science, College of Liberal Arts, Tsuda University,}\\
\small\textsl{2-1-1 Tsuda-machi Kodaira, Tokyo, 187-8577, Japan}
\vspace{+6pt}\\
\small
$^2$\small\textsl{
Faculty of Education, University of the Ryukyus,}\\
\small\textsl{1 Senbaru, Nishihara-cho, Nakagami-gun, Okinawa, 903-0213 Japan}
\vspace{+6pt}\\
\small
$^3$\small\textsl{
Kanazawa Institute of Technology,}\\
\small\textsl{7-1 Ohgigaoka, Nonoichi-shi, Ishikawa, 921-8501, Japan}
}

\date{\today}

\begin{document}

\maketitle

\begin{abstract}
Two long-standing open conjectures on planar graphs,
the Albertson-Berman conjecture and the Matheson-Tarjan conjecture,
have been extensively studied by many researchers.

\begin{itemize}
\item[{\bf (AB)}]
Every planar graph of order $n$ has an induced forest of order at least $\frac{n}{2}$.
\item[{\bf (MT)}]
Every plane triangulation of sufficiently large order $n$ has a dominating set of cardinality at most $\frac{n}{4}$.
\end{itemize}

Although partial progress and weaker bounds are known,
both conjectures remain unsolved.
To shed further light on them,
researchers have explored a variety of related notions and generalizations.
In this paper, we clarify relations among several of these notions, most notably connected domination and induced outerplanar subgraphs,
and investigate the corresponding open problems.
Furthermore, we construct an infinite family of plane triangulations of order $n$ whose connected domination number exceeds $\frac{n}{3}$.
This construction gives a negative answer to a question of Bradshaw et al. [\textit{SIAM J. Discrete Math.} \textbf{36} (2022) 1416–1435], who asked whether the maxleaf number of every plane triangulation of order $n$ is at least $\frac{2n}{3}$.
We also obtain new results on induced subgraphs with bounded treewidth and induced outerplanar subgraphs.

\end{abstract}

\noindent
\emph{Keywords and phrases.}
Planar graph,
Outerplanar graph,
$K_4$-Minor free graph,
Albertson-Berman conjecture,
Connected domination,

\noindent
\emph{AMS 2020 Mathematics Subject Classification.}
05C10, 05C69. 

\section{Introduction}

In graph theory,
one of the most natural and fundamental questions is:
Given a graph, how large can an induced particular subgraph be taken? 
Not only does this question represent a basic structural concern, but it also has implications in various contexts;
for example, it may be applicable to the study of forbidden induced subgraphs, such as determining how large an induced forest must be forbidden in order to guarantee a desired property.

In this section, we discuss known results and conjectures related to the above question divided into several subsections, 
while we introduce our results relevant to them.
Then we summarize the relation among those unresolved conjectures.
For the relevant definitions and terminology used throughout this paper, the reader is referred to Section~\ref{def}.

\subsection{Large induced forests in planar graphs}

The Albertson-Berman conjecture~\cite{AB79} concerning large induced forests,
which itself is of interest,
remains unsolved.
The maximun order of an induced forest in a graph $G$,
denoted by $f(G)$,
is called the \emph{induced forest number} of $G$.

\begin{Con}[Albertson and Berman~\cite{AB79}]
\label{con:AB79}
Let $G$ be a planar graph of order $n$.
Then $f(G)\ge\frac{n}{2}$.
\end{Con}

If true,
this would imply the existence of an independent set of cardinality at least~$\frac{n}{4}$ in planar graphs,
without relying on the Four-Color Theorem.
Several partial results have been obtained toward Conjecture~\ref{con:AB79}.

\begin{Thm}[Hosono~\cite{H90}]
\label{thm:Hosono}
Let $G$ be an outerplanar graph of order $n$.
Then~$f(G)\ge\frac{2n}{3}$.
\end{Thm}

\begin{Thm}[Le~\cite{Le18}]
Let $G$ be a triangle-free planar graph of order $n$.
Then~$f(G)\ge\frac{5n}{9}$.
\end{Thm}

The best general bound known for all planar graphs is due to Borodin~\cite{B79}, based on his result on acyclic 5-colorings:

\begin{Thm}[Borodin~\cite{B79}]
\label{borodinthm}
Let $G$ be a planar graph of order $n$.
Then $f(G)\ge\frac{2n}{5}$.
\end{Thm}

A related conjecture for bipartite planar graphs,
posed by Akiyama and Watanabe~\cite{AW87},
proposes a bound of $\frac{5n}{8}$,
which is also unresolved.

Another variation by Chappell
considers induced \emph{linear forests}
(which are forests consisting only of paths)
in planar graphs,
conjecturing a bound of $\frac{4n}{9}$ (cf. \cite{Pels04}).
Similarly to Conjecture~\ref{con:AB79},
if true,
this would also imply the existence of an independent set of cardinality at least $\frac{2n}{9}$.
Recent progress has been made in this line of research.
Notably, Dross, Montassier and Pinlou~\cite{DMP19} proved the following result:

\begin{Thm}[Dross et al.~\cite{DMP19}]
\label{thm:DMP}
Let $G$ be a triangle-free planar graph of order $n$ and of size $m$.
Then $G$ has an induced linear forest of order at least $\frac{9n-2m}{11}$.
\end{Thm}

Other recent developments include results on large induced subgraphs with bounded degree in outerplanar and planar graphs by D'Elia and Frati~\cite{DF25}.
In particular, they proved that
every planar graph of order $n$ has an induced subgraph with maximum degree at most 3
of order at least $\frac{5n}{13}$,
and also constructed a planar graph of order $n$ whose largest induced subgraph with maximum degree at most 3 of order $\frac{4n}{7} + O(1)$.

\subsection{Large induced forests in graphs with bounded treewidth}

Another key structural parameter that measures how close a graph is to a tree or a forest is \emph{treewidth}.
A \emph{tree decomposition} of a graph $G$ is a tree $T$ with vertices $X_1,\dots,X_n$,
where $X_i$ for each $i\in \{1,\dots,n\}$ is a subset of $V(G)$ such that
\begin{enumerate}
\item[(i)]
$\bigcup^n_{i=1} X_i = V(G)$,
\item[(ii)]
for each edge of $G$, there is an index $i$ such that 
both end vertices of the edge belong to $X_i$, and 
\item[(iii)]
if both $X_i$ and $X_j$ contain a vertex $v$, 
then all vertices $X_k$ of $T$ in the unique path between $X_i$ and $X_j$ contain $v$ as well. 
\end{enumerate}
The \emph{width} of a tree decomposition is defined as $\max_i|X_i| - 1$,
and the \emph{treewidth} of $G$, denoted by $tw(G)$, is the minimum width of any tree decomposition of $G$.
Note that $G$ is a forest if and only if $tw(G)=1$.

Intuitively,
graphs with small treewidth tend to admit large induced forests.
Indeed, Chappell and Pelsmajer~\cite{CP13} gave a lower bound on the order of induced forests in graphs with bounded treewidth.

\begin{Thm}[Chappell and Pelsmajer~\cite{CP13}]
\label{thm:CP13}
Let $k\ge 2$ and $d\ge 2$ be integers,
and $G$ be a graph of order $n$ with treewidth at most $k$.
Then the maximum order of an induced forest in $G$ having maximum degree at most $d$ is at least $\frac{2dn+2}{kd+d+1}$
unless $G\in\{ K_{1,1,3}, K_{2,3} \}$ and $k=d=2$.
\end{Thm}

More generally, we can consider both the treewidth of a given graph and that of its induced subgraphs to derive the following result,
which is the first result of ours.


\begin{thm}
\label{thm:treewidth}
Let $s$ and $t$ be positive integers,
and let $G$ be a graph with $tw(G)\le s$.
Then $G$ has an induced subgraph $H$
such that $tw(H)\le t$ and $|H|\ge\frac{t+1}{s+1}|G|$.
\end{thm}

This theorem generalizes Theorem~\ref{thm:Hosono}
and a result proved by Fertin, Godard and Raspaud~\cite{FGR02},
which states that every graph of order $n$ with treewidth at most $k$ contains an induced forest of order $\frac{2n}{k+1}$.

In Section~\ref{sec:proofthm1},
we shall prove a theorem stronger than Theorem~\ref{thm:treewidth}.
Theorem~\ref{thm:treewidth} has an important application as follows.
If a planar graph $G$ has an induced subgraph of order at least $p$ that has treewidth at most 2,
then $G$ has an induced forest of order $\frac{2p}{3}$ by Theorem~\ref{thm:Ange}.
Thus, the existence of $K_4$-minor free induced subgraph of order at least $\frac{2n}{3}$ (Conjecture~\ref{con:K_4} described in the next subsection) improves the best known lower bound of Theorem~\ref{borodinthm} to $\frac{4n}{9}$.

Very recently, 
Matsumoto and Yashima~\cite{MY2025+} have investigated large induced subgraphs with given pathwidth,
which is a path-variant of treewidth,
in outerplanar graphs.
This research is motivated by Chappell's conjecture
because every linear forest has pathwidth~1.

\subsection{Large induced subgraphs concerning outerplanarity}\label{Sec:1.3}

From a broader perspective,
although the Albertson-Berman and the Chappell conjectures (if true) would provide sharp lower bounds~(with examples given in~\cite{AW87} and~\cite{Pels04}, respectively),
one might expect to obtain larger lower bounds
by considering a broader class of graphs,
depending on the induced subgraphs under consideration.
Indeed, Theorem~\ref{thm:treewidth} serves as an example of this approach.
Thus, by focusing on the outerplanarity of graphs,
we can consider the following invariants:
the maximum order of an induced outerplanar (resp. outerplane, $K_4$-minor free) subgraph in a plane graph $G$, denoted by $s_o(G)$ (resp. $s_{o'}(G)$, $s_{K_4}(G)$),
is called the \emph{outerplanar} (resp. \emph{outerplane, $K_4$-minor free) subgraph number} of $G$.
Among the above three invariants, the following relation holds.

\begin{obs}
\label{obs:3outers}
For any plane graph $G$, $s_{o'}(G) \le s_{o}(G) \le s_{K_4}(G)$.
\end{obs}

Borradaile, Le and Sherman-Bennett~\cite{BLS17} showed that for every 2-outerplanar graph $G$, 
which is a planar graph close to an outerplanar graph,
$s_{o}(G)\ge\frac{2n}{3}$.
This together with Theorem~\ref{thm:treewidth} (or Theorem~\ref{thm:Hosono}) implies that such graphs admit an induced forest of order at least $\frac{4n}{9}$.
Based on this, they proposed the following conjecture.

\begin{Con}[Borradaile et al.~\cite{BLS17}]
\label{con:outerplanar}
Let $G$ be a planar graph of order $n$.
Then $s_{o}(G)\ge\frac{2n}{3}$.
\end{Con}
    
Motivated by their conjecture and Observation~\ref{obs:3outers},
we pose the following related conjecture.

\begin{con}
\label{con:K_4}
Let $G$ be a planar graph of order $n$.
Then $s_{K_4}(G)\ge\frac{2n}{3}$.
\end{con}

One may consider a similar problem for $s_{o'}(G)$:
for any plane graph $G$ of order $n$,
does $s_{o'}(G)\ge\frac{2n}{3}$ follow?
However, this is not true in general.
We construct a plane Eulerian triangulation $G$ of order $n$ with $s_{o'}(G) = \frac{11n}{18}+\frac{2}{3}$ in Section~\ref{sec:counterexample_cd} (see Remark~\ref{rem:final2}).
Generally, the following lower bound is known.

\begin{Thm}[Angelini et al.~\cite{AEFG16}]
\label{thm:Ange}
Let $G$ be a plane graph of order $n$.
Then $s_{o'}(G) \ge \frac{n}{2}$.
\end{Thm}

Although the above two conjectures remain open,
we establish the differences between $s_{o}(G)$ and $s_{K_4}(G)$, as well as between $s_{o'}(G)$ and $s_{o}(G)$ as follows,
which are the second results of ours and shall be proved in Sections \ref{sec:proof_k4o} and \ref{sec:proof_oo}, respectively.

\begin{thm}
\label{thm:diff_K_4-o}
Let $G$ be a plane triangulation.
Then the difference $s_{K_4}(G)-s_{o}(G)$ is arbitrarily large according to the order of $G$.
\end{thm}

\begin{thm}
\label{thm:diff_o-o'}
Let $G$ be a plane triangulation.
Then the difference $s_{o}(G)-s_{o'}(G)$ is arbitrarily large according to the order of $G$.
\end{thm}

Conversely, we also have two sharp upper bounds of those differences,
which is the third result of ours.

\begin{thm}
\label{thm:diff_K_4-o_upper}
Let $G$ be a planar graph of order $n$.
Then
$$s_{K_4}(G) - s_o(G) \le \frac{1}{5}s_{K_4}(G) \le \frac{n}{5}.$$
\end{thm}

\begin{thm}
\label{thm:diff_o-o'_upper}
Let $G$ be a planar graph of order $n$.
Then
$$s_o(G) - s_{o'}(G) \le \frac{s_o(G)-2}{3} \le \frac{n-2}{3}.$$
\end{thm}

In Sections~\ref{sec_proof_k4-o_upper} and~\ref{sec_proof_k4-o'_upper},
we show that every $K_4$-minor free graph contains a large induced outerplanar subgraph,
and that every outerplanar graph, when embedded on the plane (not necessarily as an outerplane graph), contains a large induced outerplane subgraph.
We then derive the above two theorems from these results.
Moreover, we construct graphs that attain the upper bounds given in Theorem~\ref{thm:diff_K_4-o_upper} and~\ref{thm:diff_o-o'_upper}.

\subsection{Dominations}

As described above,
the existence of large induced forests
asserts the existence of a large maximal independent set in planar graphs.
It is easy to see that
every maximal independent set is a dominating set in a graph under consideration.
In this subsection, 
we discuss relations among conjectures on large induced subgraphs 
and some on small vertex subsets such as a dominating set.

A subset $X$ of $V(G)$ is a \emph{dominating set} of $G$ if $V(G)=\bigcup_{x\in X}N_G[x]$.
The minimum cardinality of a dominating set of $G$,
denoted by $\gamma(G)$,
is called the \emph{domination number} of $G$.
As with the Albertson-Berman conjecture,
the Matheson-Tarjan conjecture~\cite{MT96} remains a prominent open problem on planar graphs. 
Matheson and Tarjan~\cite{MT96} showed a partial result and provided examples suggesting that the bound~$\frac{n}{4}$ would be tight if the conjecture were true.

\begin{Con}[Matheson and Tarjan~\cite{MT96}]
\label{con:MT96}
Let $G$ be a plane triangulation of sufficiently large order $n$.
Then $\gamma(G)\le \frac{n}{4}$.
\end{Con}

\begin{Thm}[Matheson and Tarjan~\cite{MT96}]
\label{thm:MT96}
Let $G$ be a plane triangulation of order $n$.
Then $\gamma(G)\le \frac{n}{3}$.
\end{Thm}

King and Pelsmajoer~\cite{KP10} proved that for a plane triangulation $G$ with maximum degree $6$, $\gamma(G)\le \frac{n}{4}$.
\v{S}pacapan~\cite{S20} showed that
for a plane triangulation $G$ of order at least $7$, $\gamma(G)\le \frac{17n}{53}$.
The best known general bound is due to Christiansen Rotenberg and Rutschmann~\cite{CRR2024}:

\begin{Thm}[Christiansen et al.~\cite{CRR2024}]
Let $G$ be a plane triangulation of order at least $11$.
Then $\gamma(G)\le \frac{2n}{7}$.
\end{Thm}

As with problems concerning induced subgraphs, domination has many variants.
In our work, we focus on total domination and connected domination.
A subset $X$ of $V(G)$ is a \emph{total dominating set} of $G$ if $V(G)=\bigcup_{x\in X}N_G(x)$.
The minimum cardinality of a total dominating set of $G$,
denoted by $\gamma_t(G)$,
is called the \emph{total domination number} of $G$.
A subset $X$ of $V(G)$ is a \emph{connected dominating set} of $G$ if $V(G)=\bigcup_{x\in X}N_G[x]$ and the subgraph $G[X]$ is connected.
The minimum cardinality of a connected dominating set of $G$,
denoted by $\gamma_c(G)$,
is called the \emph{connected domination number} of $G$.
Note that any connected dominating set consisting of at least two vertices is a total dominating set.
For the study on domination invariants,
see \cite{H22} for a recent survey, and \cite{DW12,HHH23,HY2013} for books.
Similarly to Conjecture~\ref{con:MT96},
various conjectures and questions have been formulated concering these concepts for planar graphs.

\begin{Con}[Claverol et al.~\cite{CGHHMMT2021}]
\label{con:total}
Let $G$ be a plane triangulation of order $n \ge 6$.
Then $\gamma_t(G)\le \frac{n}{3}$.
\end{Con}

Claverol, Garc\'{i}a, Hern\'{a}ndez, Hernando, Maureso, Mora and Tejel~\cite{CGHHMMT2021} also proved that $\gamma_t(G) \le \frac{2n}{5}$ for any near-triangulation $G$ of order $n \ge 5$, with two exceptions,
where a \emph{near-triangulation} is a plane graph in which all faces except possibly one are triangles.
In particular, every triangulation is a near-triangulation.

\begin{Q}[Bryant and Pavelescu~\cite{BP25}]
\label{Q:connected}
Is it true that $\gamma_c(G)\le \frac{n}{3}$ for all plane triangulation of order $n$?
\end{Q}

Recently, Bose, Dujmovi\'{c}, Houdrouge, Morin and Odak,~\cite{BDHMO23+} announced that for a plane triangulation $G$ of order $n \ge 3$,
$\gamma_c(G) \le \frac{10n}{21}$.
Zhuang~\cite{Z20} also proved that 
for a $K_4$-minor free graph $G$ of order $n\ge 6$ with
$k$ vertices of degree $2$,
$\gamma_c(G)\le\frac{2(n-k)}{3}$.
Question~\ref{Q:connected} is in fact equivalent to the following question concerning the existence of a spanning tree with the maximum number of leaves,
which is known as the \emph{maxleaf number}, 
due to Bradshaw, Masa\v{r}\'{i}k, Novotn\'{a} and Stacho~\cite{BMNS22}.

\begin{Q}[Bradshaw et al.~\cite{BMNS22}]
\label{Q:maxleaf}
Is the maxleaf number of any plane triangulation of order $n$ at least~$\frac{2n}{3}$?
\end{Q}

In connection with the above,
Noguchi and Zamfirescu~\cite{NZ2024} recently proved that 
any 4-connected plane triangulation of order $n$
has a spanning tree with exactly $k$ leaves for any integer $2 \le k \le \frac{n}{2}+1$.
Furthermore, they constructed a 4-connected plane triangulation of order $n$ 
which has no spanning tree with more than $\frac{2n}{3}$ leaves.

\subsection{Relations among the conjectures and a counterexample}

It is a notable observation that
Question~\ref{Q:connected} implies
that $s_{o'} \ge \frac{2}{3}|V(G)|$ for a plane triangulation $G$, and then it also implies Conjecture~\ref{con:outerplanar}:
Take a plane triangulation $G$ and a connected dominating set $S$ 
(if necessary, we redraw $G$ on the plane so that a vertex in $S$ appears on the infinite face).
Since $S$ is connected,
it must be contained within a single face of $G-S$.
Because $S$ dominates $G$,
every vertex in $G-S$ must be adjacent to a vertex in $S$,
and hence must appear on the boundary of that face.
It follows that $G-S$ is outerplane.

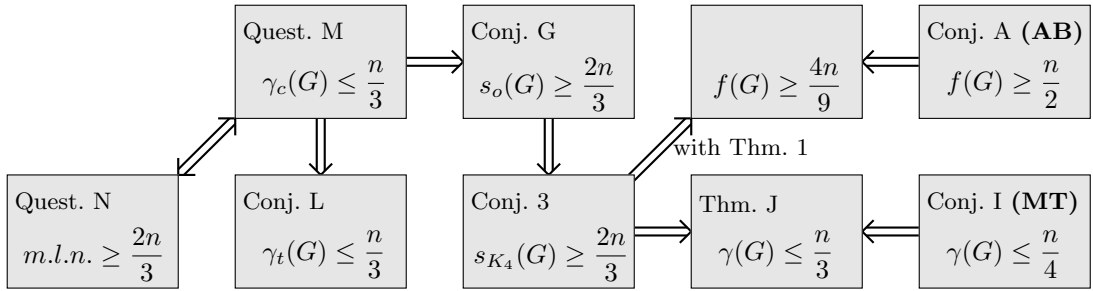
\begin{figure}[ht]
\center
\begin{center}
\begin{tikzpicture}[samples=200,scale=0.75]

\draw[thick] (-1,2+0.1)--(0-0.1,3);
\draw[thick] (-1+0.1,2)--(0,3-0.1);
\draw[thick] (-1,2+0.3)--(-1,2)--(-1+0.3,2);
\draw[thick] (0-0.3,3)--(0,3)--(0,3-0.3);
\draw[thick] (1.5-0.07,3)--(1.5-0.07,2+0.07);
\draw[thick] (1.5+0.07,3)--(1.5+0.07,2+0.07);
\draw[thick] (1.5-0.2,2+0.2)--(1.5,2)--(1.5+0.2,2+0.2);
\draw[thick] (3,4-0.07)--(4-0.07,4-0.07);
\draw[thick] (3,4+0.07)--(4-0.07,4+0.07);
\draw[thick] (4-0.2,4-0.2)--(4,4)--(4-0.2,4+0.2);
\draw[thick] (5.5-0.07,3)--(5.5-0.07,2+0.07);
\draw[thick] (5.5+0.07,3)--(5.5+0.07,2+0.07);
\draw[thick] (5.5-0.2,2+0.2)--(5.5,2)--(5.5+0.2,2+0.2);
\draw[thick] (7-0.1,2)--(8-0.1,3);
\draw[thick] (7,2-0.1)--(8,3-0.1);
\draw[thick] (8-0.3,3)--(8,3)--(8,3-0.3);
\draw[thick] (7,1+0.07)--(8-0.07,1+0.07);
\draw[thick] (7,1-0.07)--(8-0.07,1-0.07);
\draw[thick] (8-0.2,1-0.2)--(8,1)--(8-0.2,1+0.2);
\draw[thick] (11+0.07,4+0.07)--(12,4+0.07);
\draw[thick] (11+0.07,4-0.07)--(12,4-0.07);
\draw[thick] (11+0.2,4+0.2)--(11,4)--(11+0.2,4-0.2);
\draw[thick] (11+0.07,1+0.07)--(12,1+0.07);
\draw[thick] (11+0.07,1-0.07)--(12,1-0.07);
\draw[thick] (11+0.2,1+0.2)--(11,1)--(11+0.2,1-0.2);

\filldraw[white!90!black] (0-4,0)--(3-4,0)--(3-4,2)--(0-4,2)--cycle;
\filldraw[white!90!black] (0,0)--(3,0)--(3,2)--(0,2)--cycle;
\filldraw[white!90!black] (0+4,0)--(3+4,0)--(3+4,2)--(0+4,2)--cycle;
\filldraw[white!90!black] (0+8,0)--(3+8,0)--(3+8,2)--(0+8,2)--cycle;
\filldraw[white!90!black] (0+12,0)--(3+12,0)--(3+12,2)--(0+12,2)--cycle;

\filldraw[white!90!black] (0,0+3)--(3,0+3)--(3,2+3)--(0,2+3)--cycle;
\filldraw[white!90!black] (0+4,0+3)--(3+4,0+3)--(3+4,2+3)--(0+4,2+3)--cycle;
\filldraw[white!90!black] (0+8,0+3)--(3+8,0+3)--(3+8,2+3)--(0+8,2+3)--cycle;
\filldraw[white!90!black] (0+12,0+3)--(3+12,0+3)--(3+12,2+3)--(0+12,2+3)--cycle;


\draw (0-4,0)--(3-4,0)--(3-4,2)--(0-4,2)--cycle;
\draw (0,0)--(3,0)--(3,2)--(0,2)--cycle;
\draw (0+4,0)--(3+4,0)--(3+4,2)--(0+4,2)--cycle;
\draw (0+8,0)--(3+8,0)--(3+8,2)--(0+8,2)--cycle;
\draw (0+12,0)--(3+12,0)--(3+12,2)--(0+12,2)--cycle;

\draw (0,0+3)--(3,0+3)--(3,2+3)--(0,2+3)--cycle;
\draw (0+4,0+3)--(3+4,0+3)--(3+4,2+3)--(0+4,2+3)--cycle;
\draw (0+8,0+3)--(3+8,0+3)--(3+8,2+3)--(0+8,2+3)--cycle;
\draw (0+12,0+3)--(3+12,0+3)--(3+12,2+3)--(0+12,2+3)--cycle;


\draw (0-4-0.05,1.5)node[right]{\footnotesize{Quest.~\ref{Q:maxleaf}}};
\draw (0-4+0.2-0.1,0.5+0.1)node[right]{\small{$\displaystyle m.l.n.\ge\frac{2n}{3}$}};

\draw (0-0.05,1.5)node[right]{\footnotesize{Conj.~\ref{con:total}}};
\draw (0+0.4-0.1,0.5+0.1)node[right]{\small{$\displaystyle \gamma_t(G)\le\frac{n}{3}$}};

\draw (0+4-0.05,1.5)node[right]{\footnotesize{Conj.~\ref{con:K_4}}};
\draw (0+4+0.1-0.1,0.5+0.1)node[right]{\small{$\displaystyle s_{K_4}(G)\ge\frac{2n}{3}$}};

\draw (0+8-0.05,1.5)node[right]{\footnotesize{Thm.~\ref{thm:MT96}}};
\draw (0+8+0.4-0.1,0.5+0.1)node[right]{\small{$\displaystyle \gamma(G)\le\frac{n}{3}$}};

\draw (0+12-0.05,1.5)node[right]{\footnotesize{Conj.~\ref{con:MT96}~{\bf(MT)}}};
\draw (0+12+0.4-0.1,0.5+0.1)node[right]{\small{$\displaystyle \gamma(G)\le\frac{n}{4}$}};

\draw (0-0.05,1.5+3)node[right]{\footnotesize{Quest.~\ref{Q:connected}}};
\draw (0+0.4-0.1,0.5+3+0.1)node[right]{\small{$\displaystyle \gamma_c(G)\le\frac{n}{3}$}};

\draw (0+4-0.05,1.5+3)node[right]{\footnotesize{Conj.~\ref{con:outerplanar}}};
\draw (0+4+0.2-0.1,0.5+3+0.1)node[right]{\small{$\displaystyle s_{o}(G)\ge\frac{2n}{3}$}};

\draw (0+8+0.3-0.1,0.5+3+0.1)node[right]{\small{$\displaystyle f(G)\ge\frac{4n}{9}$}};

\draw (7.5,2.5) node[right]{\footnotesize{with Thm.~\ref{thm:treewidth}}};

\draw (0+12-0.05,1.5+3)node[right]{\footnotesize{Conj.~\ref{con:AB79}~{\bf(AB)}}};
\draw (0+12+0.4-0.1,0.5+3+0.1)node[right]{\small{$\displaystyle f(G)\ge\frac{n}{2}$}};

\end{tikzpicture}
\end{center}
\caption{Relations among the conjectures.}
\label{diagram}
\end{figure}

The relations among the conjectures and questions can be summarized in the following diagram~(see Figure~\ref{diagram}).
In the diagram, $G$ denotes a plane triangulation of sufficiently large order $n$.
Observe that Conjecture~\ref{con:K_4} implies Theorem~\ref{thm:MT96}:
Let $F$ be a largest induced $K_4$-minor free subgraph of $G$ and then $|F| \ge \frac{2n}{3}$.
If $V(G) \setminus V(F)$ is not a dominating set,
then there is a vertex $x \in V(F)$ such that $N_G[x] \subseteq V(F)$.
However, since $G$ is a plane triangulation, $F[N_G[x]]$ trivially contains a $K_4$-minor, a contradiction. Thus, $V(G) \setminus V(F)$ is a dominating set.

As can be seen from the diagram,
Question~\ref{Q:connected} (i.e., Question~\ref{Q:maxleaf}) is a crucial problem
whose resolution would lead to the resolution of many other conjectures.
However, we show that the answer to Question~\ref{Q:connected} is negative
by providing an infinite family of counterexamples,
which is our final result and shall be proved in Section~\ref{sec:counterexample_cd}.
It is also worth noting that the constructed counterexamples also serve as plane graphs satisfying $s_{o'}(G) = \frac{11n}{18} + \frac{2}{3}$, as discussed in Section~\ref{Sec:1.3} (see Remark~\ref{rem:final2} again).

\begin{thm}
\label{thm:counterexample}
There exists a family of plane Eulerian triangulations $G$ such that $\gamma_c(G) \ge \frac{7|V(G)|}{18} -\frac{2}{3}$.
\end{thm}

As a consequence, Question~\ref{Q:connected} can no longer be used to establish Conjecture~\ref{con:outerplanar} or any other related conjecture that would have followed from an affirmative answer.
Nevertheless, the structural connection between connected domination and induced outerplanar subgraphs remains meaningful.
Motivated by this link and our examples,
we raise the following questions:

\begin{Q}\label{Q:newQuest-connectedDomination}
    Is there a constant $c$ such that
    for every plane triangulation $G$ of order $n$,
    $\gamma_c(G) \le \frac{7n}{18}+c$.
\end{Q}

\begin{Q}\label{Q:newQuest-maxleaf}
    Is there a constant $c$ such that
    the maxleaf number of any plane triangulation of order $n$ is at least
    $\frac{11n}{18}+c$.
\end{Q}

\begin{Q}\label{Q:newQuest-Outerplane}
    Is there a constant $c$ such that
    for every plane triangulation $G$,
    $s_{o'}(G) \ge \frac{11|V(G)|}{18}+c$.
\end{Q}

We expect that the coefficients $\frac{7}{18}$ and $\frac{11}{18}$ in Questions~\ref{Q:newQuest-connectedDomination}, \ref{Q:newQuest-maxleaf} and \ref{Q:newQuest-Outerplane} are asymptotically best possible,
based on our constructions,
although the precise values of the constant terms remain undetermined.

Even if these questions turn out to have positive answers,
they do not imply Conjecture~\ref{con:outerplanar} or any related conjecture.
Nevertheless, they capture structural insights revealed by our counterexamples and remain of independent interest.
Although Question~\ref{Q:connected} fails, Conjectures~\ref{con:outerplanar} and~\ref{con:K_4} remain open.
Thus, resolving them may lead to an improved lower bound for $f(G)$, and bring us closer to a complete resolution of Conjecture~\ref{con:AB79}.

\section{Definitions and foundations}~\label{def}

All graphs considered in this paper are finite, simple, and undirected.
Let $G$ be a graph.
We let $V(G)$ and $E(G)$ denote the set of vertices and the set of edges of $G$, respectively.
For a vertex $x$ of $G$, let $N_G(x)$, $N_G[x]$ and $d_G(x)$ denote the \emph{open neighborhood}, the \emph{closed neighborhood}, and the \emph{degree} of $x$, respectively; thus $N_G(x)=\{y\in V(G) \mid xy\in E(G)\}$, $N_G[x]=N_G(x)\cup\{x\}$ and $d_G(x)=|N_G(x)|$.
For a subset $X$ of $V(G)$, we let $G[X]$ denote the subgraph of $G$ induced by $X$.
We let $K_n$ denote the complete graph of order $n$
and let $K_{m,n}$ denote the complete bipartite graph consisting of two partite sets of order $m$ and $n$.
A vertex of degree $k$ is said to be a \emph{$k$-vertex},
and a cycle of length $k$ is said to be a \emph{$k$-cycle}.
A graph $G$ is said to be \emph{maximal} with respect to a graph property $\mathcal{P}$
if it satisfies $\mathcal{P}$, and adding any edge to $G$ results in a graph that no longer satisfies $\mathcal{P}$.

A graph is said to be \emph{planar}
if it can be drawn on the sphere (or the plane) without edges crossing.
Such a drawing is called an \emph{embedding} of the graph on the sphere.
A \emph{plane graph} is a planar graph already embedded on the sphere.
A \emph{plane triangulation} is a plane graph in which every face is bounded by a $3$-cycle.
Equivalently, it can be seen as an embedding of a maximal planar graph on the sphere.
A graph is said to be \emph{outerplanar}
if it admits an embedding on the sphere in which all vertices lie on the boundary of some face.
When such an embedding is fixed,
the resulting embedded graph is called an \emph{outerplane graph},
in analogy with the notion of a plane graph.
A maximal outerplanar graph can be embedded on the sphere so that all faces, except for one face whose boundary includes all the vertices, are bounded by $3$-cycles. 

A \emph{minor} of a graph $G$ is a graph that can be obtained from $G$ by removing vertices and removing or contracting edges.
For a graph $H$, a graph $G$ is \emph{$H$-minor free} if no minor of $G$ is isomorphic to $H$.
It is well known that planar graphs are exactly the graphs that are both $K_5$-minor free and $K_{3,3}$-minor free, by Kuratowski’s and Wagner’s theorems.
In addition, outerplanar graphs are precisely the graphs that are both $K_4$-minor free and $K_{2,3}$-minor free.

For an integer $k \ge 1$,
a \emph{$k$-tree} is recursively defined as follows:
A complete graph $K_{k+1}$ is a $k$-tree.
Any graph obtained from a $k$-tree $G$ by adding a new vertex $v$ and making it adjacent to exactly $k$ vertices of $G$, such that they together $v$ induce $K_{k+1}$, is also a $k$-tree.
It is known that a graph is maximal with respect to tree-width at most $k$ if and only if it is a $k$-tree.
For $k=2$, these graphs coincide with the maximal $K_4$-minor free graphs.
Moreover, every maximal outerplanar graph is a $2$-tree, and hence a maximal $K_4$-minor free graph.

\section{Proof of Theorem~\ref{thm:treewidth}}\label{sec:proofthm1}

In this section,
we prove the following theorem stronger than Theorem~\ref{thm:treewidth}.

\begin{thm}
\label{th:ktree}
Every $k$-tree $G$ has the unique partition of $V(G)$ into $k+1$ independent sets for any integer $k \ge 1$.
Moreover, for any integer $2\leq t \leq k+1$, the union of any $t$ subsets in this partition induces a $(t-1)$-tree.
\end{thm}

\proof{} 
Let $G$ be a $k$-tree for an integer $k \ge 1$.
Let $K_{k+1} = G_0, G_1, \dots, G_m = G$
be a sequence of $k$-trees such that 
for each $0 \le i \le m-1$,
$G_{i+1}$ is obtained from $G_i$ 
by adding a new vertex and joining it to exactly $k$ vertices of a $K_k$ in $G_i$.
We first show by induction on $m$ that $G$ has the unique partition of $V(G)$ into $k+1$ independent sets.
When $m=0$, the claim trivially holds. So suppose that $m \ge 1$.
Let $v$ be a vertex of $G$ that does not belong to $G_{m-1}$ and $S$ be the set of $k$ vertices adjacent to $v$ in $G$.
By the induction hypothesis,
$G_{m-1}$ has the unique partition $U'_1,U'_2,\dots,U'_{k+1}$ of $V(G_{m-1})$ into independent sets.
Without loss of generality,
we may suppose that $|S \cap U'_i|=1$ for each $1 \le i \le k$
and $S \cap U'_{k+1} = \emptyset$.
Let $U_i=U'_i$ for $1 \le i \le k$ and $U_{k+1}=U'_{k+1}\cup\{v\}$.
Then $U_1,U_2,\dots,U_{k+1}$ forms a partition of $V(G)$ into independent sets.
Suppose that $G$ admits another such partition $U_1^*,U_2^*,\dots,U_{k+1}^*$ of $V(G)$.
Without loss of generality, we may assume that $v \in U_{k+1}^*$.
Then removing $v$ from $U_{k+1}^*$ yields a partition of $V(G_{m-1})$ into $k+1$ independent sets,
which must coincide with $U'_1,U'_2,\dots,U'_{k+1}$.
This implies that $U_i^*=U_i$ for all $1\le i\le k+1$, a contradiction.
Therefore, $U_1,U_2,\dots,U_{k+1}$ is the unique partition of $V(G)$ into independent sets.



Next, we show that for any integer $2\leq t \leq k+1$, the union of any $t$ subsets in the unique partition induces a $(t-1)$-tree $H$.
Let $U_1,U_2,\dots,U_{k+1}$ be the unique partition of $V(G)$ into independent sets.
It suffices to prove that $G'=G[V(G) \setminus U_{k+1}]$ is a $(k-1)$-tree.
Let us recall the process of constructing $G$ as a sequence $K_{k+1} = G_0, G_1, \dots, G_m = G$.
Let $G'_0, G'_1, \dots, G'_m=G'$ be the process of constructing $G'$
obtained from the above process by deleting the vertices in $U_{k+1}$,
that is, $G'_i = G_i[V(G'_i)\setminus U_{k+1}]$ for $1\le i\le m$.
Note that $G'_0 = K_k$.
Let $v_{i+1}$ be the vertex of $G_{i+i}$ that does not belong to $G_i$.
If $v_{i+1} \in U_j$ for $j \in \{1,2,\dots,k\}$,
then $v$ is adjacent to exactly $k-1$ vertices in $G'_{i}$ in which they together with $v$ form a $K_{k}$.
If $v_{i+1}\in U_{k+1}$,
then we have $G'_{i+1} = G'_{i}$.
Thus, every graph generated in the process $K_k = G'_0, G'_1, \dots, G'_m = G'$ is a $(k-1)$-tree,
although the number of distinct construction steps may be less than $m$ since $G'_{i+1} = G'_i$ may occur.
\qed

Recall that a $k$-tree is a maximal graph with respect to treewidth at most $k$,
i.e., 
every graph with treewidth at most $k$ is a subgraph of a $k$-tree.
By Theorem~\ref{th:ktree},
if a graph $G$ has treewidth at most $s$,
$G$ has a partition of $V(G)$ into $s+1$ independent sets $U_1,U_2,\dots,U_{s+1}$
with $|U_1| \ge |U_2| \ge \cdots \ge |U_{s+1}|$
(where some of them are possibly empty).
Any $t+1$ subsets in the partition induce a graph with treewidth at most $t$ by Theorem~\ref{th:ktree}.
Therefore, an induced subgraph $H = G[U_1 \cup U_2 \cup \cdots \cup U_{t+1}]$
has treewidth at most $t$ and its order is at least $\frac{(t+1)|V(G)|}{s+1}$,
and then Theorem~\ref{thm:treewidth} follows.

\section{Proof of Theorem~\ref{thm:diff_K_4-o}}\label{sec:proof_k4o}

We show that the difference $s_{K_4}(G)-s_{o}(G)$ is arbitrarily large according to the order of a plane triangulation $G$.
To see this,
we construct a plane triangulation $D_k$ 
such that $s_{K_4}(D_k) \ge \frac{3(|V(D_k)|-1)}{4}$ and $s_{o}(D_k) \le \frac{23(|V(D_k)|-1)}{32}$.

\begin{figure}[ht]
\center
\begin{center}
\begin{tikzpicture}[scale=3]

\coordinate (u1) at ({cos(90)}, {sin(90)});
\coordinate (u2) at ({cos(210)}, {sin(210)});
\coordinate (u3) at ({cos(330)}, {sin(330)});
\coordinate (u1') at ({1.03*cos(90)}, {1.03*sin(90)});
\coordinate (u2') at ({1.03*cos(210)}, {1.03*sin(210)});
\coordinate (u3') at ({1.03*cos(330)}, {1.03*sin(330)});

\draw (u1')node[above]{$a$};
\draw (u2')node[below]{$b$};
\draw (u3')node[below]{$c$};

\def\scaleFactorOne{0.2}

\coordinate (v1) at ({1.5*\scaleFactorOne*cos(150)}, {1.5*\scaleFactorOne*sin(150)});
\coordinate (v2) at ({1.5*\scaleFactorOne*cos(270)}, {1.5*\scaleFactorOne*sin(270)});
\coordinate (v3) at ({1.5*\scaleFactorOne*cos(390)}, {1.5*\scaleFactorOne*sin(390)});
\coordinate (v1') at ({1.55*\scaleFactorOne*cos(150)}, {1.55*\scaleFactorOne*sin(150)});
\coordinate (v2') at ({1.55*\scaleFactorOne*cos(270)}, {1.55*\scaleFactorOne*sin(270)});
\coordinate (v3') at ({1.55*\scaleFactorOne*cos(390)}, {1.55*\scaleFactorOne*sin(390)});

\def\scaleFactorTwo{0}

\coordinate (w1) at ({1.7*\scaleFactorTwo*cos(90)}, {1.3*\scaleFactorTwo*sin(90)});
\coordinate (w1') at (0,0.02);
\draw (w1')node[above]{$v$};

\foreach\P in{u1,u2,u3}\fill[black](\P)circle(0.04);
\foreach\P in{v1,v2,v3}\fill[black](\P)circle(0.04);
\foreach\P in{w1}\fill[black](\P)circle(0.04);

\draw[semithick] (u1) to (u2);
\draw[semithick] (u2) to (u3);
\draw[semithick] (u3) to (u1);

\draw[semithick] (v1) to (v2);
\draw[semithick] (v2) to (v3);
\draw[semithick] (v3) to (v1);

\draw[semithick](v3)--(u1)--(v1);
\draw[semithick](v1)--(u2)--(v2);
\draw[semithick](v2)--(u3)--(v3);

\draw[semithick](v1)--(w1);
\draw[semithick](v2)--(w1);
\draw[semithick](v3)--(w1);

\end{tikzpicture}
\end{center}
\caption{The graph $R'$.}
\label{fig:thm5-1}
\end{figure}
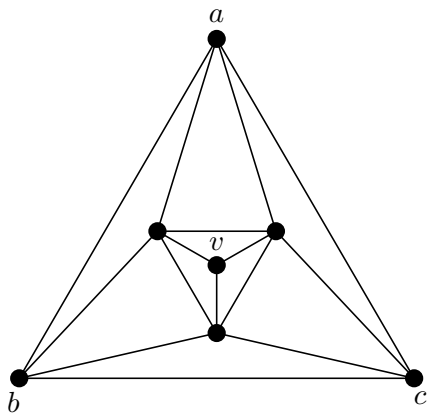

\bigskip
\noindent
{\bf Construction:}
We first prepare 
a plane embedding $O$ of the octahedron graph
and insert a single vertex $v$ into a face of $O$ joined to all the vertices on the boundary of the face.
The resulting graph is denoted by $R'$ and the boundary of the face consisting of all $4$-vertices is denoted by $abc$ (see Figure~\ref{fig:thm5-1}).
Next we prepare a $K_4$ on the plane, where $V(K_4)=\{x_0,x_1,x_2,x_3\}$.
For each integer $ 1\le i \le 3$, let $F_i$ denote the face bounded by $x_i, x_{i+1}$ and $x_0$, where the indices are taken modulo $3$.
We replace $F_i$ with a copy of $R'$, identifying the boundary $3$-cycle $x_0x_{i}x_{i+1}$ of $F_i$ with the boundary cycle $abc$ of $R'$.
The resulting graph is denoted by $R$.
Note that $R$ can be partitioned into exactly four $K_4$s. 

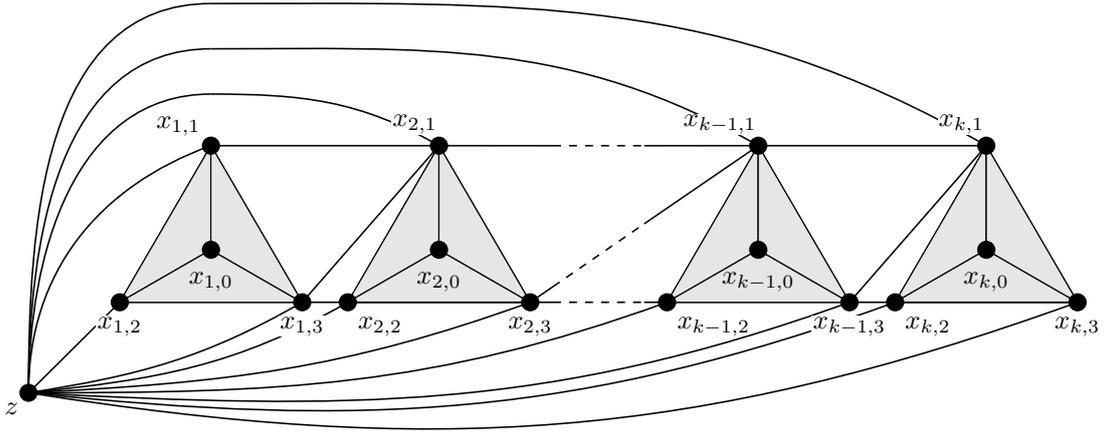
\begin{figure}[ht]
\center
\begin{center}
\begin{tikzpicture}[scale=1.2]

\coordinate (z) at (-1,-1);
\foreach\P in{z}\fill[black](\P)circle(0.1);

\coordinate (u1) at (0,0); 
\coordinate (v1) at (2,0); 
\coordinate (w1) at (1,1.73); 
\coordinate (z1) at (1,0.58); 
\coordinate (u1') at (0,0-0.05);
\coordinate (v1') at (2,0-0.05);
\coordinate (z1') at (1,0.45);

\draw[semithick](u1)--(z1)--(w1)--cycle;
\filldraw[fill=gray!20!white](u1)--(z1)--(w1)--cycle;

\draw[semithick](u1)--(v1)--(z1)--cycle;
\filldraw[fill=gray!20!white](u1)--(v1)--(z1)--cycle;

\draw[semithick](v1)--(w1)--(z1)--cycle;
\filldraw[fill=gray!20!white](v1)--(w1)--(z1)--cycle;

\foreach\P in{u1,v1,w1,z1}\fill[black](\P)circle(0.1);

\coordinate (u2) at (0+2.5,0);
\coordinate (v2) at (2+2.5,0);
\coordinate (w2) at (1+2.5,1.73);
\coordinate (z2) at (1+2.5,0.58);
\coordinate (u2') at (0+2.5,0-0.05);
\coordinate (v2') at (2+2.5,0-0.05);
\coordinate (z2') at (1+2.5,0.45);

\draw[semithick](u2)--(z2)--(w2)--cycle;
\filldraw[fill=gray!20!white](u2)--(z2)--(w2)--cycle;

\draw[semithick](u2)--(v2)--(z2)--cycle;
\filldraw[fill=gray!20!white](u2)--(v2)--(z2)--cycle;

\draw[semithick](v2)--(w2)--(z2)--cycle;
\filldraw[fill=gray!20!white](v2)--(w2)--(z2)--cycle;

\foreach\P in{u2,v2,w2,z2}\fill[black](\P)circle(0.1);

\coordinate (u3) at (0+6,0);
\coordinate (v3) at (2+6,0);
\coordinate (w3) at (1+6,1.73);
\coordinate (z3) at (1+6,0.58);
\coordinate (u3') at (0+6,0-0.05);
\coordinate (v3') at (2+6,0-0.15);
\coordinate (z3') at (1+6,0.45);'

\draw[semithick](u3)--(z3)--(w3)--cycle;
\filldraw[fill=gray!20!white](u3)--(z3)--(w3)--cycle;

\draw[semithick](u3)--(v3)--(z3)--cycle;
\filldraw[fill=gray!20!white](u3)--(v3)--(z3)--cycle;

\draw[semithick](v3)--(w3)--(z3)--cycle;
\filldraw[fill=gray!20!white](v3)--(w3)--(z3)--cycle;

\foreach\P in{u3,v3,w3,z3}\fill[black](\P)circle(0.1);

\coordinate (u4) at (0+8.5,0);
\coordinate (v4) at (2+8.5,0);
\coordinate (w4) at (1+8.5,1.73);
\coordinate (z4) at (1+8.5,0.58);
\coordinate (u4') at (0+8.5,0-0.05);
\coordinate (v4') at (2+8.5,0-0.05);
\coordinate (z4') at (1+8.5,0.45);

\draw[semithick](u4)--(z4)--(w4)--cycle;
\filldraw[fill=gray!20!white](u4)--(z4)--(w4)--cycle;

\draw[semithick](u4)--(v4)--(z4)--cycle;
\filldraw[fill=gray!20!white](u4)--(v4)--(z4)--cycle;

\draw[semithick](v4)--(w4)--(z4)--cycle;
\filldraw[fill=gray!20!white](v4)--(w4)--(z4)--cycle;

\foreach\P in{u4,v4,w4,z4}\fill[black](\P)circle(0.1);

\coordinate (a) at (1,2.3);
\coordinate (b) at (1,2.8);
\coordinate (c) at (1,3.3);
\coordinate (d) at (4.5+0.25,0.173);
\coordinate (e) at (4.5+1.25,0.867);

\draw[semithick](w1)--(w2);
\draw[semithick](v1)--(w2);
\draw[semithick](v1)--(v2);

\draw[semithick](w2)--(4.75,1.73);
\draw[semithick,dashed](4.75,1.73)--(5.75,1.73);
\draw[semithick](5.75,1.73)--(w3);

\draw[semithick](v2)--(d);
\draw[semithick,dashed](d)--(e);
\draw[semithick](e)--(w3);

\draw[semithick](v2)--(4.75,0);
\draw[semithick,dashed](4.75,0)--(5.75,0);
\draw[semithick](5.75,0)--(u3);

\draw[semithick](w3)--(w4);
\draw[semithick](v3)--(w4);
\draw[semithick](v3)--(v4);

\draw[semithick] (z) to [out=90,in=180] (a);
\draw[semithick] (z) to [out=90,in=180] (b);
\draw[semithick] (z) to [out=90,in=180] (c);

\draw[semithick] (a) to [out=0,in=150] (w2);
\draw[semithick] (b) to [out=0,in=150] (w3);
\draw[semithick] (c) to [out=0,in=150] (w4);

\draw[semithick] (z) to [out=90,in=200] (w1);
\draw[semithick] (z) to (u1);
\draw[semithick] (z) to [out=9,in=210] (v1);
\draw[semithick] (z) to [out=6,in=210] (u2);
\draw[semithick] (z) to [out=3,in=200] (v2);
\draw[semithick] (z) to [out=0,in=200] (u3);
\draw[semithick] (z) to [out=-3,in=200] (v3);
\draw[semithick] (z) to [out=-6,in=200] (u4);
\draw[semithick] (z) to [out=-9,in=200] (v4);

\draw (z)node[below left]{{\small$z$}}; 

\draw (0,0-0.15)node[below, fill=white, inner xsep=0.5pt, inner ysep=0.2pt]{{\small$x_{1,2}$}}; 
\draw (2,0-0.15)node[below, fill=white, inner xsep=0.5pt, inner ysep=0.2pt]{{\small$x_{1,3}$}}; 
\draw (w1)node[above left]{{\small$x_{1,1}$}}; 
\draw (z1')node[below]{{\small$x_{1,0}$}}; 

\draw (u2')node[below right]{{\small$x_{2,2}$}}; 
\draw (v2')node[below]{{\small$x_{2,3}$}}; 
\draw (3.5,1.85)node[above left, fill=white, inner xsep=0.5pt, inner ysep=0.2pt]{{\small$x_{2,1}$}}; 
\draw (z2')node[below]{{\small$x_{2,0}$}}; 

\draw (u3')node[below right]{{\small$x_{k-1,2}$}}; 
\draw (v3')node[below, fill=white, inner xsep=0.5pt, inner ysep=0.2pt]{{\small$x_{k-1,3}$}}; 
\draw (7,1.85)node[above left, fill=white, inner xsep=0.5pt, inner ysep=0.2pt]{{\small$x_{k-1,1}$}}; 
\draw (z3')node[below]{{\small$x_{k-1,0}$}}; 

\draw (u4')node[below right]{{\small$x_{k,2}$}}; 
\draw (v4')node[below]{{\small$x_{k,3}$}}; 
\draw (9.5,1.85)node[above left, fill=white, inner xsep=0.5pt, inner ysep=0.2pt]{{\small$x_{k,1}$}}; 
\draw (z4')node[below]{{\small$x_{k,0}$}}; 

\end{tikzpicture}
\end{center}
\caption{The graph $D_k$, where a copy of $R'$ is embedded into each shaded triangle.}
\label{fig:thm5-2}
\end{figure}

Let $k$ be an even integer with $k \ge 4$,
and let $R_1, \dots, R_k$ be disjoint copies of $R$.
For each integer $1 \le j \le k$,
let $x_{j,0}, x_{j,1}, x_{j,2}$ and $x_{j,3}$ 
be the vertices of $R_j$ corresponding to $x_0,x_1,x_2$ and $x_3$ of $R$, respectively.
We now define the graph $D_k$ (see Figure~\ref{fig:thm5-2}) by
\begin{align*}
V(D_k)
&=\{z\}  \cup  \bigcup_{1 \le j \le k}V(R_j), \text{ where all } R_j \text{ are mutually disjoint, and} \\
E(D_k)
&=\{ x_{i,1}x_{i+1,1}, x_{i,3}x_{i+1,1}, x_{i,3}x_{i+1,2} \mid 1 \le i \le k-1 \} \\
&\quad \cup \{xz \,|\, x \in \bigcup_{1 \le j \le k}\{x_{j,1},x_{j,2},x_{j,3}\}\} \cup \bigcup_{1 \le j \le k}E(R_j).
\end{align*}
\medskip

We first consider $s_{o}(D_k)$.
Let $H$ be a largest induced outerplanar subgraph of $D_k$.
Since every outerplanar graph is $K_4$-minor free,
$H$ takes at most three vertices of each $K_4$ subgraph of $D_k$,
that is, $|V(H) \cap V(R_j)| \le 3 \cdot 4 = 12$ for each integer $1\le j \le k$.
Moreover, if $|V(H) \cap V(R_j)|=12$,
then $x_{j,1},x_{j,2},x_{j,3} \in V(H) \cap V(R_j)$;
otherwise, for example,
if $x_{j,1},x_{j,2},x_{j,0} \in V(H) \cap V(R_j)$,
then they together with three vertices in the triangular region
$x_{j,1}x_{j,2}x_{j,0}$ belonging to $V(H) \cap V(R_j)$ contains a $K_4$-minor.
Moreover,
if there exists $ i \in \{ 1,\ldots, k-1\}$
such that $|V(H) \cap V(R_i)|=12$ and $|V(H) \cap V(R_{i+1})|=12$,
$\{x_{i,1},x_{i,2},x_{i,3},x_{i+1,1},x_{i+1,2},x_{i+1,3}\}$
and three vertices of $H$ in the interior of the region bounded by $x_{i,0}x_{i,1}x_{i,3}$ induce a $K_{2,3}$-minor.
Thus, for each odd integer $1 \le j \le k-1$,
$|V(H) \cap V(R_j)|+|V(H) \cap V(R_{j+1})| \leq 12 + 11 = 23$.
It is easy to see that $z$ does not belong to $V(H)$ if there is an index $j$ such that $|V(H) \cap V(R_j)|=12$.
Since $|V(D_k)|=16k+1$,
we have
$$s_{o}(D_k) = |V(H)| \le 23 \cdot \frac{k}{2} = \frac{23k}{2} = \frac{23(|V(D_k)|-1)}{32}.$$

Next, we consider $s_{K_4}(D_k)$.
We can easily see that
$M = D_k[\bigcup_{1 \le j \le k}\{x_{j,1},x_{j,2},x_{j,3}\}]$ is a maximal outerplanar subgraph,
that is, a $2$-tree.
For each copy of $R'$, 
by construction,
we may assume that the vertex corresponding to $a$ does not belong to $V(M)$,
while the vertices corresponding to $b$ and $c$ do belong to $V(M)$.
Furthermore, among the four vertices in the copy of $R'$ that correspond to the vertices other than $a$, $b$ and $c$,
we can choose three such vertices
so that the subgraph induced by these three vertices together with $V(M)$ remains a $2$-tree.
Since the three chosen vertices in one copy of $R'$ are not adjacent to those chosen in any other copy,
we can obtain a $K_4$-minor free induced subgraph with $12k=\frac{3(|V(D_k)|-1)}{4}$ vertices in total from $D_k$.

Therefore, 
the difference $s_{K_4}(D_k) - s_o(D_k)$ is arbitrarily large according to the order of the triangulation.
This completes the proof of Theorem~\ref{thm:diff_K_4-o}.

\section{Proof of Theorem~\ref{thm:diff_o-o'}}\label{sec:proof_oo}

We show that the difference $s_o(G)-s_{o'}(G)$ is arbitrarily large according to the order of a plane triangulation $G$.
To see this,
we construct 
a plane triangulation $G_k$
such that $s_{o}(G_k) \ge \frac{15}{22}|V(G_k)|$
and such that $s_{o'}(G_k) \le \frac{7}{11}|V(G_k)|$.

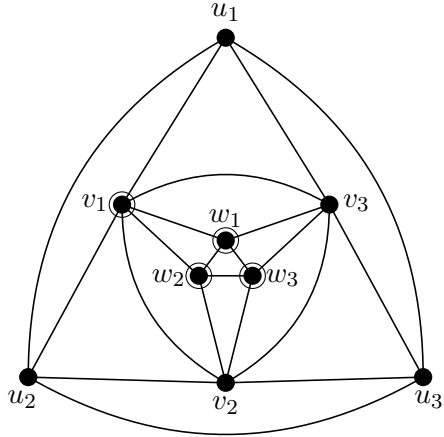
\begin{figure}[ht]
\center
\begin{center}
\begin{tikzpicture}[scale=3]


\coordinate (u1) at ({cos(90)}, {sin(90)});
\coordinate (u2) at ({cos(210)}, {sin(210)});
\coordinate (u3) at ({cos(330)}, {sin(330)});
\coordinate (u1') at ({1.03*cos(90)}, {1.03*sin(90)});
\coordinate (u2') at ({1.03*cos(210)}, {1.03*sin(210)});
\coordinate (u3') at ({1.03*cos(330)}, {1.03*sin(330)});


\draw (u1')node[above]{$u_1$};
\draw (u2')node[below]{$u_2$};
\draw (u3')node[below]{$u_3$};

\def\scaleFactorOne{0.35}

\coordinate (v1) at ({1.5*\scaleFactorOne*cos(150)}, {1.5*\scaleFactorOne*sin(150)});
\coordinate (v2) at ({1.5*\scaleFactorOne*cos(270)}, {1.5*\scaleFactorOne*sin(270)});
\coordinate (v3) at ({1.5*\scaleFactorOne*cos(390)}, {1.5*\scaleFactorOne*sin(390)});
\coordinate (v1') at ({1.55*\scaleFactorOne*cos(150)}, {1.55*\scaleFactorOne*sin(150)});
\coordinate (v2') at ({1.55*\scaleFactorOne*cos(270)}, {1.55*\scaleFactorOne*sin(270)});
\coordinate (v3') at ({1.55*\scaleFactorOne*cos(390)}, {1.55*\scaleFactorOne*sin(390)});

\draw (v1')node[left]{$v_1$};
\draw (v2')node[below]{$v_2$};
\draw (v3')node[right]{$v_3$};

\def\scaleFactorTwo{0.08}

\coordinate (w1) at ({1.7*\scaleFactorTwo*cos(90)}, {1.3*\scaleFactorTwo*sin(90)});
\coordinate (w2) at ({1.7*\scaleFactorTwo*cos(210)}, {1.3*\scaleFactorTwo*sin(210)});
\coordinate (w3) at ({1.7*\scaleFactorTwo*cos(330)}, {1.3*\scaleFactorTwo*sin(330)});
\coordinate (w1') at ({1.9*\scaleFactorTwo*cos(90)}, {1.5*\scaleFactorTwo*sin(90)});
\coordinate (w2') at ({1.9*\scaleFactorTwo*cos(210)}, {1.5*\scaleFactorTwo*sin(210)});
\coordinate (w3') at ({1.9*\scaleFactorTwo*cos(330)}, {1.5*\scaleFactorTwo*sin(330)});

\draw (w1')node[above]{$w_1$};
\draw (w2')node[left]{$w_2$};
\draw (w3')node[right]{$w_3$};

\foreach\P in{u1,u2,u3}\fill[black](\P)circle(0.04);
\foreach\P in{v1,v2,v3}\fill[black](\P)circle(0.04);
\foreach\P in{w1,w2,w3}\fill[black](\P)circle(0.04);

\draw[semithick] (u1) to [out=210,in=90] (u2);
\draw[semithick] (u2) to [out=330,in=210] (u3);
\draw[semithick] (u3) to [out=90,in=330] (u1);

\draw[semithick] (v1) to [out=270,in=150] (v2);
\draw[semithick] (v2) to [out=30,in=270] (v3);
\draw[semithick] (v3) to [out=150,in=30] (v1);

\draw[semithick](w1)--(w2)--(w3)--cycle;

\draw[semithick](v3)--(u1)--(v1);
\draw[semithick](v1)--(u2)--(v2);
\draw[semithick](v2)--(u3)--(v3);

\draw[semithick](v1)--(w1)--(v3);
\draw[semithick](v1)--(w2)--(v2);
\draw[semithick](v2)--(w3)--(v3);

\draw (v1) circle[radius=1.6pt];
\draw (w1) circle[radius=1.6pt];
\draw (w2) circle[radius=1.6pt];
\draw (w3) circle[radius=1.6pt];

\end{tikzpicture}
\end{center}
\caption{The graph $I$, where the four vertices enclosed in circles form the set $S$.}
\label{fig1}
\end{figure}

\medskip
\noindent
{\bf Construction:} 
For any integer $k \ge 4$,
we construct the graph $G_k$ as follows.
We first prepare
a plane embedding $O$ of the octahedron graph
and replace an inner face $F$ of $O$
with a copy of another octahedron $O'$
by identifying the $3$-cycle of $F$
with the boundary cycle of $O'$.
The resulting graph is denoted by $I$.
Moreover, set $S=\{ v_1, w_1, w_2, w_3 \}$~(see Figure~\ref{fig1}).

We next prepare a $K_4$ with vertex set $V(K_4)=\{x_0,x_1,x_2,x_3\}$, embedded on the plane so that the outer face is bounded by the $3$-cycle $x_1x_2x_3$.
For each integer $1 \le i \le 3$, let $F_i$ denote the face bounded by $x_i, x_{i+1}$ and $x_0$, where the indices are taken modulo $3$.
We replace $F_i$ with a copy of $I$, identifying the boundary $3$-cycle of $F_i$ with the boundary cycle $u_1 u_2 u_3$ of $I$.
The resulting graph is denoted by $H$.

\begin{figure}[ht]
\center
\begin{center}
\begin{tikzpicture}[scale=1.3]

\coordinate (u1) at (0,0);
\coordinate (v1) at (2,0);
\coordinate (w1) at (1,1.73);
\coordinate (z1) at (1,0.58);
\coordinate (u1') at (0,0-0.05);
\coordinate (v1') at (2,0-0.05);
\coordinate (z1') at (1,0.45);

\draw[semithick](u1)--(z1)--(w1)--cycle;
\filldraw[fill=gray!20!white](u1)--(z1)--(w1)--cycle;

\draw[semithick](u1)--(v1)--(z1)--cycle;
\filldraw[fill=gray!20!white](u1)--(v1)--(z1)--cycle;

\draw[semithick](v1)--(w1)--(z1)--cycle;
\filldraw[fill=gray!20!white](v1)--(w1)--(z1)--cycle;

\foreach\P in{u1,v1,w1,z1}\fill[black](\P)circle(0.1);

\coordinate (u2) at (0+2.5,0);
\coordinate (v2) at (2+2.5,0);
\coordinate (w2) at (1+2.5,1.73);
\coordinate (z2) at (1+2.5,0.58);
\coordinate (u2') at (0+2.5,0-0.05);
\coordinate (v2') at (2+2.5,0-0.05);
\coordinate (z2') at (1+2.5,0.45);

\draw[semithick](u2)--(z2)--(w2)--cycle;
\filldraw[fill=gray!20!white](u2)--(z2)--(w2)--cycle;

\draw[semithick](u2)--(v2)--(z2)--cycle;
\filldraw[fill=gray!20!white](u2)--(v2)--(z2)--cycle;

\draw[semithick](v2)--(w2)--(z2)--cycle;
\filldraw[fill=gray!20!white](v2)--(w2)--(z2)--cycle;

\foreach\P in{u2,v2,w2,z2}\fill[black](\P)circle(0.1);

\coordinate (u3) at (0+6,0);
\coordinate (v3) at (2+6,0);
\coordinate (w3) at (1+6,1.73);
\coordinate (z3) at (1+6,0.58);
\coordinate (u3') at (0+6,0-0.05);
\coordinate (v3') at (2+6,0-0.15);
\coordinate (z3') at (1+6,0.45);'

\draw[semithick](u3)--(z3)--(w3)--cycle;
\filldraw[fill=gray!20!white](u3)--(z3)--(w3)--cycle;

\draw[semithick](u3)--(v3)--(z3)--cycle;
\filldraw[fill=gray!20!white](u3)--(v3)--(z3)--cycle;

\draw[semithick](v3)--(w3)--(z3)--cycle;
\filldraw[fill=gray!20!white](v3)--(w3)--(z3)--cycle;

\foreach\P in{u3,v3,w3,z3}\fill[black](\P)circle(0.1);

\coordinate (u4) at (0+8.5,0);
\coordinate (v4) at (2+8.5,0);
\coordinate (w4) at (1+8.5,1.73);
\coordinate (z4) at (1+8.5,0.58);
\coordinate (u4') at (0+8.5,0-0.05);
\coordinate (v4') at (2+8.5,0-0.05);
\coordinate (z4') at (1+8.5,0.45);

\draw[semithick](u4)--(z4)--(w4)--cycle;
\filldraw[fill=gray!20!white](u4)--(z4)--(w4)--cycle;

\draw[semithick](u4)--(v4)--(z4)--cycle;
\filldraw[fill=gray!20!white](u4)--(v4)--(z4)--cycle;

\draw[semithick](v4)--(w4)--(z4)--cycle;
\filldraw[fill=gray!20!white](v4)--(w4)--(z4)--cycle;

\foreach\P in{u4,v4,w4,z4}\fill[black](\P)circle(0.1);

\coordinate (a) at (1+8.5-0.6,1.73+0.5);
\coordinate (b) at (1+8.5-0.75,1.73+1);
\coordinate (c) at (1+8.5-0.9,1.73+1.5);
\coordinate (d) at (4.5+0.25,0.173);
\coordinate (e) at (4.5+1.25,0.867);

\draw[semithick](w1)--(w2);
\draw[semithick](v1)--(w2);
\draw[semithick](v1)--(v2);

\draw[semithick](w2)--(4.75,1.73);
\draw[semithick,dashed](4.75,1.73)--(5.75,1.73);
\draw[semithick](5.75,1.73)--(w3);

\draw[semithick](v2)--(d);
\draw[semithick,dashed](d)--(e);
\draw[semithick](e)--(w3);

\draw[semithick](v2)--(4.75,0);
\draw[semithick,dashed](4.75,0)--(5.75,0);
\draw[semithick](5.75,0)--(u3);

\draw[semithick](w3)--(w4);
\draw[semithick](v3)--(w4);
\draw[semithick](v3)--(v4);

\draw[semithick] (v1) to [out=-30,in=210] (v2);
\draw[semithick] (u1) to [out=-30,in=210] (v2);
\draw[semithick] (v2) to [out=-30,in=210] (v3);
\draw[semithick] (v2) to [out=-30,in=210] (u3);
\draw[semithick] (v2) to [out=-30,in=210] (u4);
\draw[semithick] (v2) to [out=-30,in=210] (v4);

\draw[semithick] (u1) to [out=-30,in=210] (v4);

\draw[semithick] (w3) to [out=30,in=180] (a);
\draw[semithick] (w2) to [out=30,in=180] (b);
\draw[semithick] (w1) to [out=30,in=180] (c);

\draw[semithick] (a) to [out=0,in=100] (v4);
\draw[semithick] (b) to [out=0,in=90] (v4);
\draw[semithick] (c) to [out=0,in=80] (v4);

\draw (u1')node[below]{{\small$x_{1,2}$}}; 
\draw (v1')node[below]{{\small$x_{1,3}$}}; 
\draw (w1)node[above left]{{\small$x_{1,1}$}}; 
\draw (z1')node[below]{{\small$x_{1,0}$}}; 

\draw (u2')node[below right]{{\small$x_{2,2}$}}; 
\draw (v2')node[below]{{\small$x_{2,3}$}}; 
\draw (w2)node[above left]{{\small$x_{2,1}$}}; 
\draw (z2')node[below]{{\small$x_{2,0}$}}; 

\draw (u3')node[below right]{{\small$x_{k-1,2}$}}; 
\draw (v3')node[below, fill=white, inner xsep=0.5pt, inner ysep=0.2pt]{{\small$x_{k-1,3}$}}; 
\draw (w3)node[above left]{{\small$x_{k-1,1}$}}; 
\draw (z3')node[below]{{\small$x_{k-1,0}$}}; 

\draw (u4')node[below right]{{\small$x_{k,2}$}}; 
\draw (v4')node[below]{{\small$x_{k,3}$}}; 
\draw (w4)node[above left]{{\small$x_{k,1}$}}; 
\draw (z4')node[below]{{\small$x_{k,0}$}}; 

\end{tikzpicture}
\end{center}
\caption{The graph $G_k$, where a copy of $I$ is embedded into each shaded triangle.}
\label{fig2}
\end{figure}
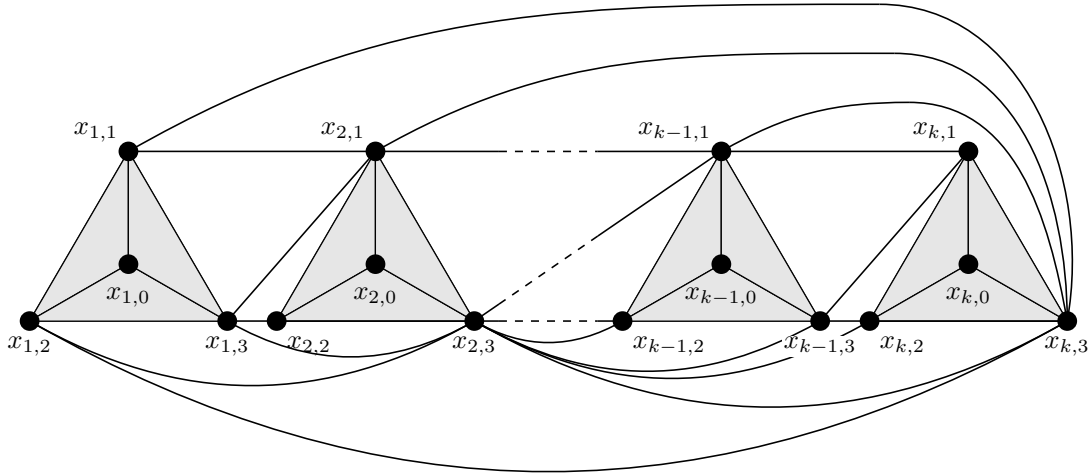

Let $k$ be an integer with $k \ge 4$,
and let $H_1, \dots, H_k$ be disjoint copies of $H$.
For each integer $1 \le j \le k$,
let $x_{j,1}, x_{j,2}, x_{j,3}$ and $x_{j,0}$
be the vertices of $V(H_j)$
which correspond to $x_1, x_2, x_3$ and $x_0$,
respectively.
We now define the graph $G_k$ (see Figure~\ref{fig2}) by
\begin{align*}
V(G_k)
&=\bigcup_{1 \le j \le k}V(H_j),  \text{ where all } H_j \text{ are mutually disjoint, and}\\
E(G_k)
&=\{ x_{i,1}x_{i+1,1}, x_{i,3}x_{i+1,1}, x_{i,3}x_{i+1,2}, x_{i,1}x_{k,3} \mid 1 \le i \le k-1 \}\\
& \hspace{0.5cm} \cup \{ x_{2,3}x_{i,2}, x_{2,3}x_{i,3} \mid 3 \le i \le k \}
\cup \{ x_{1,2}x_{k,3}, x_{1,2}x_{2,3}, x_{1,3}x_{2,3} \}
\cup \bigcup_{1 \le j \le k}E(H_j).
\end{align*}

We first consider $s_{o'}(G_k)$.
Let $Q$ be a largest induced outerplane subgraph of $G_k$.
Similarly to the proof of Theorem~\ref{thm:diff_K_4-o},
$Q$ takes at most four vertices from an octahedron graph of $I$.
Moreover, $Q$ takes at most two vertices can be selected from each $K_4$ of $G_k$:
otherwise, for example, 
if $Q$ takes three vertices $x_{j,1},x_{j,2},x_{j,3}$,
then $Q$ cannot take any vertex from the octahedron graph in the interior of the region bounded by $x_{j,1}x_{j,2}x_{j,3}$, which contradicts to the maximality of $Q$.
Therefore, we have $s_{o'}(G_k) \le (4 \cdot 3 + 2)k = 14k = \frac{7}{11}|V(G_k)|$;
note that $|V(G_k)|=22k$.

Next, we consider $s_{o}(G_k)$.
For each integer $1 \le i \le 3$ and each integer $1 \le j \le k$,
let $S_{j,i}$ denote the subset of $V(H_j)$ which corresponds to $S$ in $F_{j,i}$,
where $F_{j,i}$ denotes a region of $H_j$ correponding to $F_i$,
and set $X_j=\{ x_{j,1}, x_{j,2}, x_{j,0} \}$.
Now the induced subgraph 
$$R=G_k \left[\left(\bigcup_{1 \le j \le k, 1 \le i \le 3}S_{j,i}\right) \cup \left(\bigcup_{1 \le j \le k}X_j\right) \right]$$
is clearly outerplanar;
note that for any pair of $i,j$ with $i \ne j$,
there is only one edge $x_{i,1}x_{j,1}$ between $H_i$ and $H_j$.
Therefore, we have $s_{o}(G_k) \ge (4 \cdot 3 + 3)k = 15k = \frac{15}{22}|V(G_k)|$.

Hence, the graph $G_k$ shows that the difference $s_o(G_k)-s_{o'}(G_k)$ is arbitrarily large according to the order.
This completes the proof of Theorem~\ref{thm:diff_o-o'}.

\section{Proof of Theorem~\ref{thm:diff_K_4-o_upper}}\label{sec_proof_k4-o_upper}

To prove Theorem~\ref{thm:diff_K_4-o_upper},
we give a sharp lower bound on $s_o(G)$ in terms of $s_{K_4}(G)$.

\begin{thm}
\label{th:ratioK4minorfreeToOuterplanr}
Let $G$ be a planar graph.
Then, $\frac{4}{5}s_{K_4}(G) \leq s_o(G)$.
\end{thm}

Note that 
we can immediately obtain Theorem~\ref{thm:diff_K_4-o_upper}
by Theorem~\ref{th:ratioK4minorfreeToOuterplanr}.
There exist graphs that attain this bound,
which are introduced at the end of this section.

To prove Theorem~\ref{th:ratioK4minorfreeToOuterplanr},
it suffices to show the following lemma.

\begin{lem}
\label{lem:KeyLemmaKO}
Any $K_4$-minor free graph $G$ of order $n$ has an induced outerplanar subgraph of order at least $\frac{4n}{5}$.
\end{lem}

Before proving Lemma~\ref{lem:KeyLemmaKO},
we prepare some definitions and notations.
For a positive integer $k$,
a \emph{$k$-clique} in a graph $G$ is a clique of order $k$ in $G$.
For a $2$-tree $G$,
we define a graph $\Delta_G$ as follows:
The vertex set of $\Delta_G$ consists of the $3$-cliques in $G$.
Two vertices of $\Delta_G$ are adjacent
if they share a common pair of adjacent vertices of $G$.
Note that every spanning tree of $\Delta_G$ can be regarded as a tree decomposition of $G$.
If $G$ is (maximal) outerplanar,
then $\Delta_G$ is a tree with maximum degree at most three.
Moreover, if $G$ is already embedded on the plane as an outerplane graph,
then $\Delta_G$ corresponds to the graph obtained from the dual of $G$
by removing the vertex corresponding to the outerface of $G$.

The following lemma clearly holds.

\begin{lem}
\label{lem:block}
If every block of a graph $G$ is outerplanar, then $G$ is also outerplanar.
\end{lem}

Outerplanar graphs can be characterized as graphs that contain neither $K_4$ nor $K_{2,3}$ as minors,
while for $2$-trees,
outerplanarity can be determined by
whether $K_{1,1,3}$ is contained as a subgraph or not,
where $K_{1,1,3}$ denotes the graph obtained from $K_2$ by adding three independent vertices and joining them to all vertices of the $K_2$.

\begin{lem}
\label{lem:K23in2tree}
A $2$-tree $G$ is not maximal outerplanar if and only if
$G$ has a subgraph isomorphic to $K_{1,1,3}$.
\end{lem}

\proof{ of Lemma~\ref{lem:K23in2tree}}
It is clear that if a graph $G$ has $K_{1,1,3}$ as a subgraph,
then $G$ is not outerplanar.
Hence, we shall show the necessity.
We suppose that a $2$-tree $G$ has no $K_{1,1,3}$ as a subgraph.
Then, each edge of $G$ is contained in at most two triangles.
We shall show that $G$ is a maximal outerplanar graph by induction on the order of $G$.
Since every $2$-tree of order at most $4$ is outerplanar, 
we may assume that $G$ has at least $5$ vertices.
Then, there exists a $2$-vertex $a$ in $G$.
Let $N_G(a)=\{b,c\}$.
Let $G'$ be the graph obtained from $G$ by removing $a$.
Then $G'$ is also a $2$-tree and the edge $bc$ belongs to at most one triangle in $G'$.
By induction hypothesis, $G'$ can be embedded on the plane
so that all vertices of $G'$ appear on the boundary of the outerface.
In this embedding, the edge $bc$ belongs to the boundary of the outerface.
In this setting, adding the vertex $a$ on the outerface of $G'$ and adding edges $ab$ and $ac$, we can obtain an embedding of $G$, which is an outerplane graph.
\qed

For a rooted tree $(T,r)$,
if a vertex $v$ in $(T,r)$ has no child,
then we call $v$ a \emph{leaf},
and if $v$ has at least one child and all its children are leaf,
then we call $v$ a \emph{pre-leaf}.

\proof{ of Lemma~\ref{lem:KeyLemmaKO}}
Let $G$ be a counterexample to Lemma~\ref{lem:KeyLemmaKO} with the smallest number of vertices.
We may assume that $G$ is a $2$-tree.
Non-outerplanar $2$-tree of order at most $5$ must be $K_{1,1,3}$,
which has an induced outerplanar subgraph of order $4$.
Thus, we may assume that the order of $G$ is at least $6$.

\begin{claim}
\label{cl:atMostThreeleaves}
For any vertex $v$ of $G$,
the neighborhood of $v$ contains at most three $2$-vertices.
\end{claim}
\proof{}
Suppose that $N_G(v)$ contains four $2$-vertices $u_1,u_2,u_3$ and $u_4$.
Let $G'=G-\{ v,u_1,u_2,u_3,u_4 \}$.
By the minimality of $G$,
there exists an induced outerplanar subgraph $H'$ of order at least $\frac{4}{5}\vert V(G')\vert$ in $G'$.
The induced subgraph $H=G[V(H')\cup \{ u_1,u_2,u_3,u_4 \}]$ can be obtained from $H'$ by adding pendant vertices $u_1,u_2,u_3$ and $u_4$.
Hence, $H$ is outerplanar and we have
$$\vert V(H) \vert = \vert V(H') \vert + 4 \geq \frac{4}{5} \vert V(G') \vert + 4 = \frac{4}{5}(\vert V(G') \vert + 5) = \frac{4}{5}\vert V(G) \vert,$$
which is a contradiction.
\qed

We choose a spanning tree $T$ of $\Delta_G$ and a root $R$
so that the number of leaves is maximized.
Note that we have $\vert V(T)\vert \geq 4$ since $\vert V(G)\vert \geq 6$.
Let $P=\{a,b,c\}$ be a pre-leaf of $(T,R)$.

\begin{claim}
\label{cl:atMostFourleavs}
Every vertex in $(T,R)$ has at most four leaves as children.
\end{claim}
\proof{}
If a vertex $Q$ in $(T,R)$ has at least five leaves as children,
then at least one of the vertices in the $3$-clique $Q$ is adjacent to at least four $2$-vertices in $G$, which contradicts Claim~\ref{cl:atMostThreeleaves}.
\qed

\begin{figure}[ht]
\center
\includegraphics[scale=0.3]{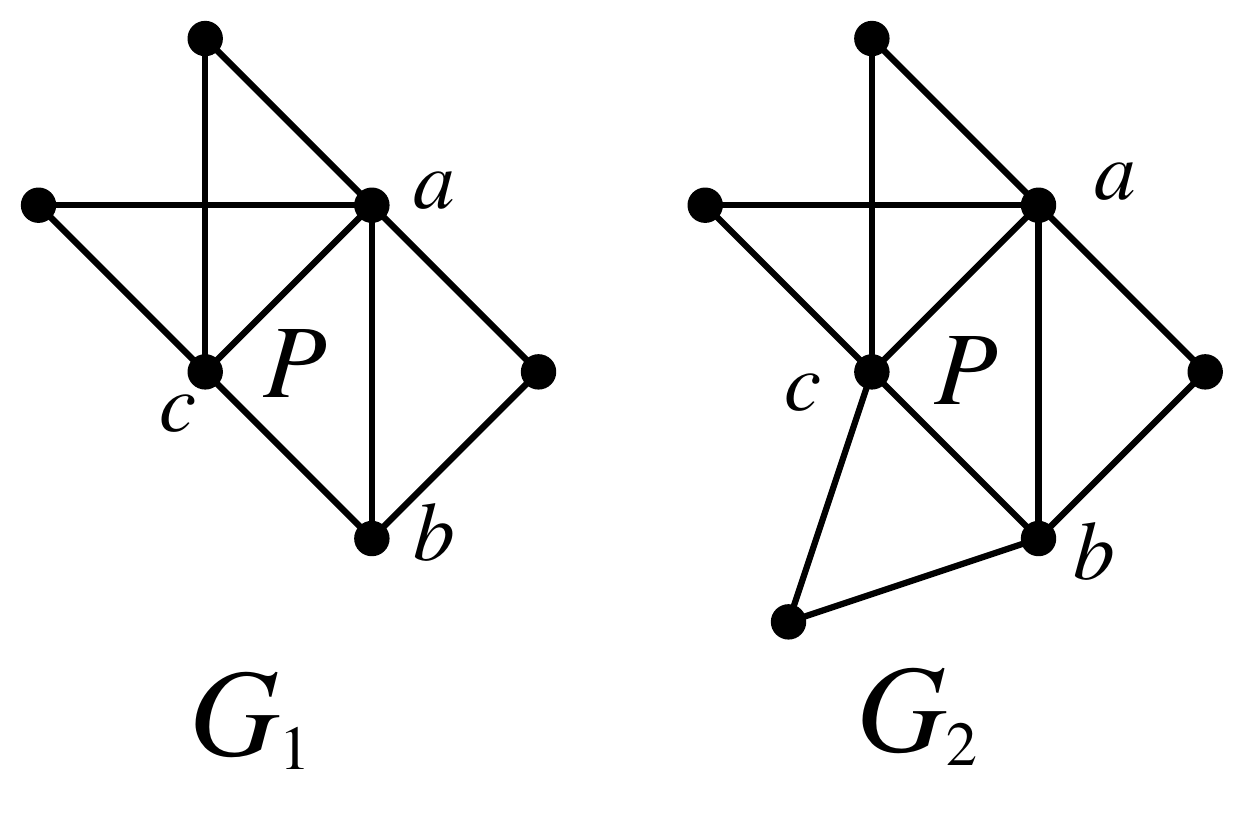}
\caption{Two graphs $G_1$ and $G_2$.}
\label{fig:G1andG2}
\end{figure}

\begin{claim}
\label{cl:notRoot}
$P$ is not the root $R$.
\end{claim}
\proof{}
Suppose that $P$ is the root $R$.
It follows from Claim~\ref{cl:atMostFourleavs} that
$P$ has at most four leaves as children,
and hence $\vert V(T) \vert \leq 5$, that is, $\vert V(G) \vert \leq 7$.
As $\vert V(G) \vert \geq 6$,
$G$ is isomorphic to $G_1$ or $G_2$ as shown in Figure~\ref{fig:G1andG2}.
However, $G_1$ and $G_2$ have induced outerplanar subgraphs of order $5$ and $6$, respectively, a contradiction. \qed

\begin{claim}
\label{cl:nonLeafHasNoDegree2}
If a $3$-clique $P'=\{v_1,v_2,v_3\}$ in $G$ is neither the root $R$ nor a leaf in $(T,R)$,
then $d_G(v_i)\geq 3$ for any $1\leq i\leq 3$.
\end{claim}
\proof{}
Let $P'=\{v_1,v_2,v_3\}$ be a vertex in $(T,R)$ which is neither the root nor a leaf.
Suppose that $d_G(v_1)=2$.
Then, the parent and every child of $P'$ include $\{v_2,v_3\}$,
that is, the parent and every child of $P'$ are adjacent in $\Delta_G$.
For each children $C$ of $P'$,
removing the edge $P'C$ and
adding the edge between $C$ and the parent of $P'$,
we can obtain another spanning tree $T'$ of $\Delta_G$.
That has one more leaf than $T$,
which contradicts the choice of $(T,R)$.
\qed

Let $Q$ be the parent of the pre-leaf $P=\{a,b,c\}$.
We may suppose that both $P$ and $Q$ include $\{a,b\}$,
and let $d$ be the vertex in the $3$-clique $Q$ but not in $P$;
Thus, $Q=\{a,b,d\}$.

\begin{claim}
\label{cl:degreeC}
$3\leq d_G(c)\leq 5$.
\end{claim}
\proof{}
Since $P$ is pre-leaf, it follows from Claim~\ref{cl:nonLeafHasNoDegree2} that $d_G(c)\geq 3$.
We suppose that $d_G(c) \geq 6$.
Since every neighbor of $c$ other than $a$ and $b$ has degree $2$,
$c$ is adjacent to at least four $2$-vertices,
which contradicts Claim~\ref{cl:atMostThreeleaves}.
\qed

\begin{figure}[ht]
\center
\includegraphics[scale=0.4]{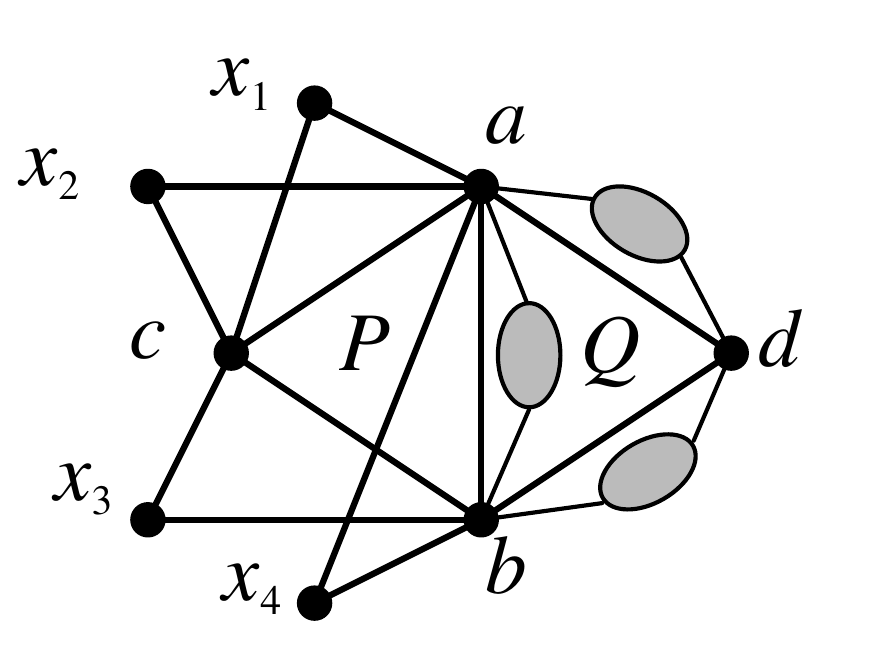}
\caption{The structure of $G$.}
\label{fig:aroundP}
\end{figure}

Let $X$ be the set of $2$-vertices of $G$
whose neighborhood is included in $P$
and that is not $d$.
Let $X$ consist of $k$ vertices $x_1,x_2,\ldots , x_k$.
Then, $P$ has $k$ children.
Since $P$ is pre-leaf, it follows from Claim~\ref{cl:atMostFourleavs} that $1\leq k\leq 4$.
By Claim~\ref{cl:degreeC}, we have $d_G(c)\geq 3$, and hence
there is a vertex in $X$ adjacent to $c$.
Thus, we may assume that $N_G(x_1)=\{a,c\}$,
that is, a $3$-clique $\{a,c,x_1\}$ is a child of $P$.
In this setting, $G$ has the structure shown in Figure~\ref{fig:aroundP},
where shaded areas represent subgraphs induced by the vertices that are not contained in $P\cup Q \cup X$.
(Some of the vertices $x_2,x_3$ and $x_4$ in Figure~\ref{fig:aroundP} may not exist,
and their neighbors may be different.)

\begin{claim}
\label{cl:twoChildren}
There are exactly two two children of $P$, that is, $k=2$,
and $N_G(x_1)=N_G(x_2)=\{a,c\}$.
\end{claim}
\proof{}
First, suppose that there are only two $3$-cliques including $\{a,c\}$:
$P$ and $\{a,c,x_1\}$.
Then, there is no subgraph having $x_1$ and isomorphic to $K_{1,1,3}$.
Let $G'=G-x_1$.
By the minimality of $G$,
there exists an induced outerplanar subgraph $H'$ of order at least $\frac{4}{5}\vert V(G')\vert$ in $G'$.
The induced subgraph $H=G[V(H')\cup \{ x_1 \}]$ has no $K_{1,1,3}$ as a subgraph, and hence $H$ is outerplanar by Lemma~\ref{lem:K23in2tree},
while $\vert V(H)\vert = \vert V(H')\vert + 1 > \frac{4}{5}(\vert V(G') \vert +1) = \frac{4}{5}\vert V(G) \vert$, a contradiction.
Therefore, $k\geq 2$ and we may assume that $P$ has a child $\{a,c,x_2\}$, that is, $N_G(x_2)=\{a,c\}$.

Second, we suppose that $k\geq 3$.
It follows from Claim~\ref{cl:atMostThreeleaves} that
there is at most two $3$-cliques including $\{b,c\}$,
one of which is $P$.
Then, there is no subgraph having the edge $bc$ and isomorphic to $K_{1,1,3}$.
Let $G'=G-(X\cup\{a,c\})$.
By the minimality of $G$,
there exists an induced outerplanar subgraph $H'$ of order at least $\frac{4}{5}\vert V(G')\vert$ in $G'$.
The induced subgraph $H=G[V(H')\cup X \cup\{c\}]$ has no $K_{1,1,3}$ as a subgraph
and hence $H$ is outerplanar by Lemma~\ref{lem:K23in2tree}.
However, since $k\geq 3$, we have
$$\vert V(H) \vert = \vert V(H') \vert + k+1
\geq \frac{4}{5}\left(\vert V(G')\vert +k+\frac{1}{4}k+\frac{5}{4}\right)
\geq \frac{4}{5}\left(\vert V(G')\vert +k+2\right)
\geq \frac{4}{5}\vert V(G) \vert ,$$
which is a contradiction.
\qed

Since every pre-leaf in $(T,R)$ is not the root by Claim~\ref{cl:notRoot},
there exists a vertex of $(T,R)$ which is the parent of a pre-leaf and
whose every child is either a leaf or a pre-leaf.
Hereafter, we suppose that $Q$ is such a vertex.

\begin{claim}
\label{cl:noChild}
$Q$ has no leaf as a child.
\end{claim}
\proof{}
Suppose that $Q$ has a leaf as a child.
Then, there exists a $2$-vertex $y$ of $G$ whose neighborhood is included in $Q$.
Let $Y=\{c,x_1,x_2,y\}$,
and if $N_G(y)$ contains $a$, we let $Z=Y\cup \{a\}$, and otherwise let $Z=Y\cup \{b\}$.
Let $G'=G-Z$.
By the minimality of $G$,
there exists an induced outerplanar subgraph $H'$ of order at least $\frac{4}{5}\vert V(G')\vert$ in $G'$.
It is easy to see that the induced subgraph $H=G[V(H')\cup Y]$ has no $K_{1,1,3}$ as a subgraph and hence $H$ is outerplanar by Lemma~\ref{lem:K23in2tree}.
However, we have $\vert V(H) \vert = \vert V(H') \vert + 4 \geq \frac{4}{5}\vert V(G) \vert$, which is a contradiction.
\qed

By Claim~\ref{cl:noChild}, all children of $Q$ are pre-leaves.

\begin{figure}[ht]
\center
\begin{tikzpicture}[scale=0.6, every node/.style={font=\Large}]

    \node[draw, circle, fill=black, inner sep=2pt, label=right:$a$] (a) at (4,4) {};
    \node[draw, circle, fill=black, inner sep=2pt, label=right:$b$] (b) at (4,0) {};
    \node[draw, circle, fill=black, inner sep=2pt, label=below left:$c$] (c) at (1,2) {};
    \node[draw, circle, fill=black, inner sep=2pt, label=right:$d$] (d) at (7,2) {};
    \node[draw, circle, fill=black, inner sep=2pt, label=left:$x_1$] (x1) at (2,4.5) {};
    \node[draw, circle, fill=black, inner sep=2pt, label=left:$x_2$] (x2) at (0.8,3.5) {};

    \draw[line width=1.5pt] (a) -- (b);
    \draw[line width=1.5pt] (a) -- (c);
    \draw[line width=1.5pt] (a) -- (d);
    \draw[line width=1.5pt] (b) -- (c);
    \draw[line width=1.5pt] (b) -- (d);
    \draw[line width=1.5pt] (x1) -- (a);
    \draw[line width=1.5pt] (x1) -- (c);
    \draw[line width=1.5pt] (x2) -- (a);
    \draw[line width=1.5pt] (x2) -- (c);

    \node at (3, 2.1) {\huge $P$};
    \node at (5, 2) {\huge $Q$};

\end{tikzpicture}
\caption{A $2$-tree of order $6$.}
\label{fig:graphOfOrder6}
\end{figure}
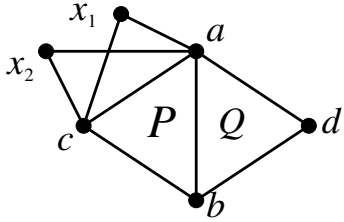

\begin{claim}
\label{cl:QHasTwoChildren}
$Q$ has at least two children.
\end{claim}
\proof{}
Suppose that $Q$ has only one child $P$.
If $Q$ is the root $R$,
then by Claim~\ref{cl:twoChildren}, $G$ is isomorphic to the graph as shown in Figure~\ref{fig:graphOfOrder6},
which has an induced outerplanar subgraph of order $5$, a contradiction.
Thus, we may assume that $Q$ is not the root $R$.

Let $Q'$ be the parent of $Q$.
It follows from Claim~\ref{cl:nonLeafHasNoDegree2} that $Q'$ contains $d$,
since otherwise $d_G(d)$ would be $2$.
Let $Z=\{a,b,c,x_1,x_2\}$,
and if $Q\cap Q'=\{a,d\}$, we let $Y=\{b,c,x_1,x_2\}$, and if $Q\cap Q'=\{b,d\}$ let $Y=\{a,c,x_1,x_2\}$.
Let $G'=G-Z$.
By the minimality of $G$,
there exists an induced outerplanar subgraph $H'$ of order at least $\frac{4}{5}\vert V(G')\vert$ in $G'$.
It is easy to see that the induced subgraph $H=G[V(H')\cup Y]$ has no $K_{1,1,3}$ as a subgraph, and hence $H$ is outerplanar by Lemma~\ref{lem:K23in2tree}.
However, we have $\vert V(H) \vert = \vert V(H') \vert + 4 \geq \frac{4}{5}\vert V(G) \vert$, which is a contradiction.
\qed

\begin{figure}[ht]
\center
\includegraphics[scale=0.35]{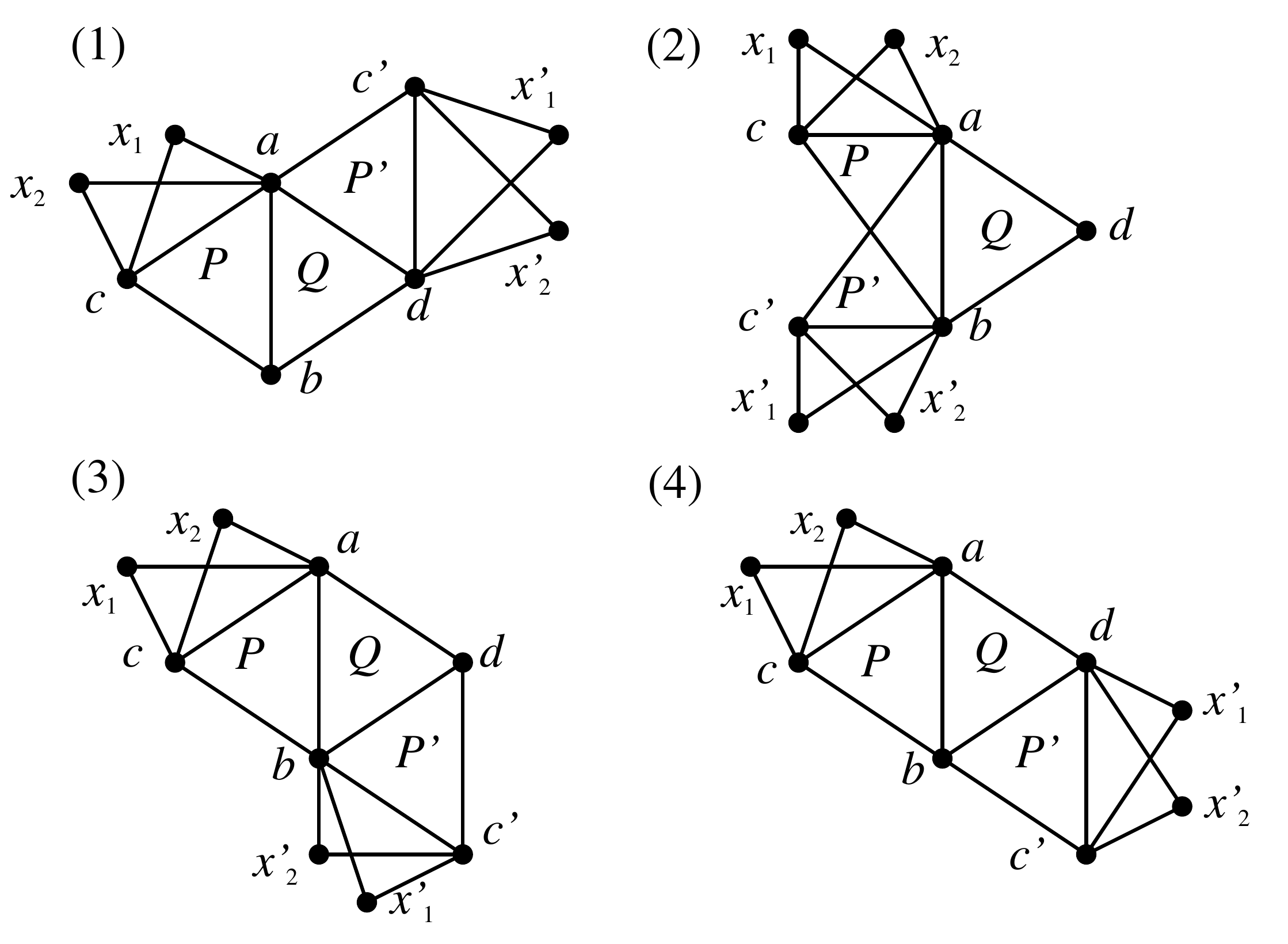}
\caption{The four possible situations around $P$ and $P'$.}
\label{fig:4type}
\end{figure}

Let $P'$ be a child of $Q$ which is not $P$,
and $c'$ be the vertex in $P'$ but not in $Q$.
Similarly to $P$, we may assume that $P'$ has two children.
Then, there exists two $2$-vertices $x'_1$ and $x'_2$
such that $N_G(x'_1)=N_G(x'_2)\subset P'$.
It follows from Claim~\ref{cl:atMostThreeleaves} that
we only have to consider the following four situations: $(1)$ to $(4)$,
as shown in Figure~\ref{fig:4type}.

Let $Y=\{c,c',x_1,x_2,x'_1,x'_2\}$,
and if the case is type $(1)$, let $Z=Y\cup \{a\}$, and otherwise, let $Z=Y\cup \{b\}$.
Let $G'=G-Z$.
By the minimality of $G$,
there exists an induced outerplanar subgraph $H'$ of order at least $\frac{4}{5}\vert V(G')\vert$ in $G'$.
It is easy to see that the induced subgraph $H=G[V(H')\cup Y]$ has no $K_{1,1,3}$ as a subgraph,
and hence $H$ is outerplanar by Lemma~\ref{lem:K23in2tree}.
However, we have
$\vert V(H) \vert = \vert V(H') \vert + 6 > \frac{4}{5}(\vert V(G') \vert + 5) = \frac{4}{5}\vert V(G) \vert$,
which is a contradiction.
Therefore, the proof of Lemma~\ref{lem:KeyLemmaKO} is finished.
\qed

\begin{figure}[ht]
\center
\includegraphics[scale=0.35]{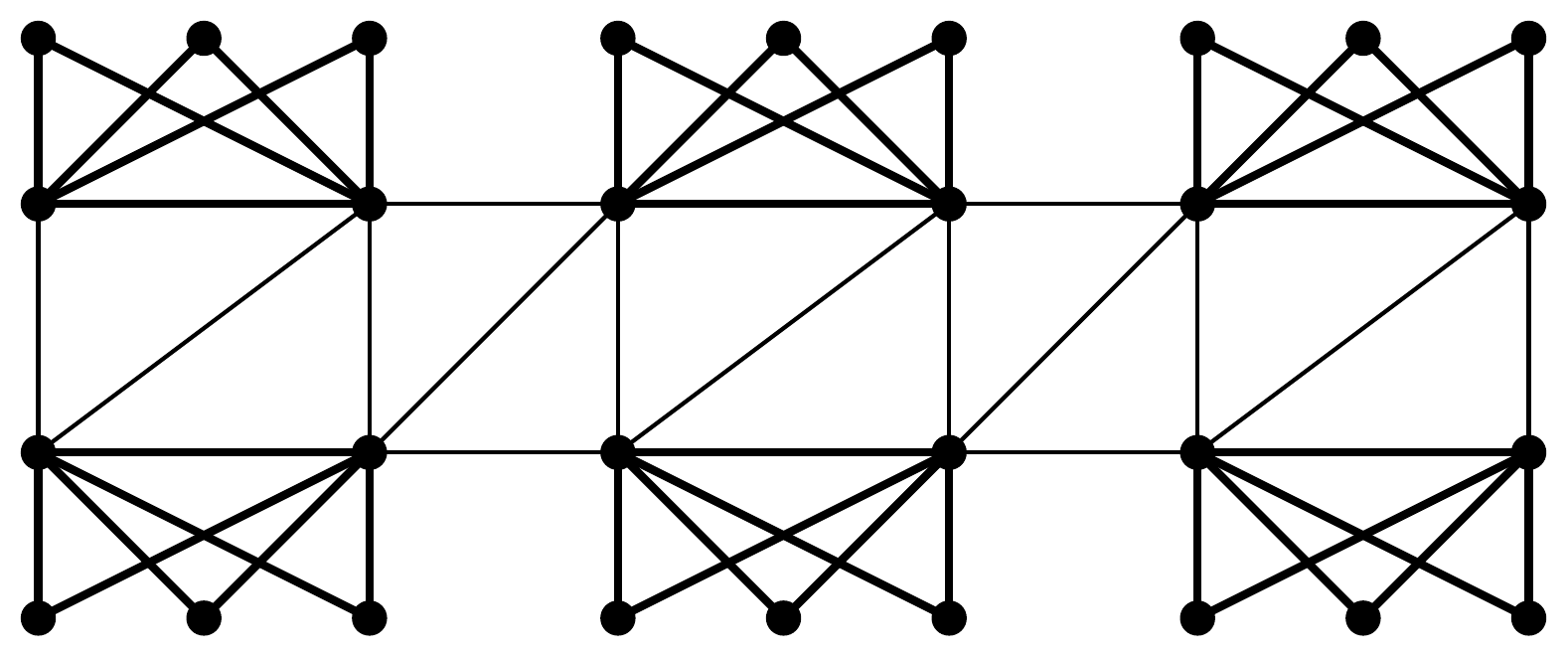}
\caption{A $2$-tree attaining the bound given by Theorem~\ref{th:ratioK4minorfreeToOuterplanr}.}
\label{fig:tight2tree}
\end{figure}

Now we construct a family of $K_4$-minor free graphs that shows that the bound given by Theorem~\ref{th:ratioK4minorfreeToOuterplanr} is tight.
We take several disjoint copies of $K_{1,1,3}$ (or $K_{2,3}$) and add any number of edges between them,
as long as the resulting graph remains $K_4$-minor free.
For example, a graph shown in Figure~\ref{fig:tight2tree} is such a construction.

\section{Proof of Theorem~\ref{thm:diff_o-o'_upper}}\label{sec_proof_k4-o'_upper}

To prove Theorem~\ref{thm:diff_o-o'_upper},
it is sufficient to prove the following.

\begin{thm}
\label{th:OuterplaneSubgraphIn2tree}
Let $G$ be a $K_4$-minor free graph of order $n$.
For any embedding of $G$ on the plane,
$G$ has an induced outerplane subgraph of order at least $\lceil\frac{2n+2}{3}\rceil$.
\end{thm}

This bound can even be attained by outerplanar graphs,
which is introduced after proving Theorem~\ref{th:OuterplaneSubgraphIn2tree}.
It follows from Theorem~\ref{th:OuterplaneSubgraphIn2tree} and its example that the following corollary provides the best possible bound,
which leads to Theorem~\ref{thm:diff_o-o'_upper}.

\begin{cor}
Let $G$ be a plane graph of order $n$.
Then,
$$ \left\lceil \frac{2s_{K_4}(G)+2}{3}\right\rceil \leq \left\lceil \frac{2s_{o}(G)+2}{3}\right\rceil \leq  s_{o'}(G).$$
\end{cor}

Let $H$ be a plane graph. 
A cycle $C$ in $H$ is \emph{separating}
if there exist vertices $p$ and $q$ of $H$ such that $p$ and $q$ lie inside $C$ and outside $C$, respectively.
In this situation, we say that $C$ \emph{separates $p$ and $q$}.
Note that $H$ is an outerplane graph if and only if there is no separating cycle in $H$.

\proof{ of Theorem~\ref{th:OuterplaneSubgraphIn2tree}}
It suffices to verify the theorem for $2$-trees.
Let $G$ be a $2$-tree of order $n$ embedded on the plane.
The case where $G$ has at most five vertices can be easily checked.
Thus, we may assume that $G$ has at least six vertices.

By Theorem~\ref{th:ktree}, $V(G)$ can be partitioned into three independent sets $V_1, V_2$ and $V_3$ so that the union of any two of them induces a forest.
We may assume that $\vert V_1 \vert \geq \vert V_2 \vert \geq \vert V_3 \vert$.
It is easy to see that the inequality $\vert V_1 \cup V_2 \vert > \lceil\frac{2n+2}{3}\rceil$ holds if $\vert V_2 \vert > \vert V_3 \vert$.
Moreover, it also holds if $\vert V_2 \vert=\vert V_3 \vert$ and $\vert V_1 \vert \geq \vert V_2 \vert + 2$.
Since any embedding of a forest is an outerplane graph,
the induced subgraph $G[V_1\cup V_2]$ is an outerplane graph.
Thus, we may assume that $\vert V_2 \vert=\vert V_3 \vert$ and $\vert V_2 \vert \leq \vert V_1 \vert \leq \vert V_2\vert + 1$.
Let $V_3$ consist of $m$ vertices $x_1,x_2,\ldots ,x_m$.
Then, $\vert V_2\vert=m$ and $m\leq \vert V_1\vert\leq m+1$.
Note that $m\geq 2$ since $G$ has at least six vertices.
In this setting, we have $\vert V_1 \vert + \vert V_2 \vert + 1 \geq \lceil\frac{2n+2}{3}\rceil$.
Thus, we may assume that there is no vertex $x$ in $V_3$ such that the induced subgraph $G[V_1\cup V_2 \cup \{x\}]$ is an outerplane graph;
otherwise, the proof is complete.

For each integer $1\leq i \leq m$, the induced subgraph $H_i=G[V_1\cup V_2 \cup \{x_i\}]$ is not an outerplane graph,
and hence there exists a separating cycle in $H_i$.
Furthermore, we now show that such a cycle can be chosen to have length $3$.
Suppose that there exists a separating cycle $\ell'_i$ in $H_i$ whose length is at least $4$.
The vertices of $\ell'_i$ other than $x_i$ are contained in $V_1\cup V_2$.
Since $G$ is a $2$-tree,
it is chordal, and hence $\ell'_i$ has a chord $e$.
If $e$ is not incident with $x_i$,
then there exists a cycle whose vertices belong to $V_1\cup V_2$,
which contradicts the fact that $V_1\cup V_2$ induces a forest.
Thus, $e$ is incident with $x_i$.
Let $p$ and $q$ be vertices of $H_i$
such that $p$ lies inside $\ell'_i$ and $q$ lies outside $\ell'_i$.
Then, we can find a shorter cycle than $\ell'_i$ in $H_i$ which includes $e$ and separates $p$ and $q$.
Therefore, we can eventually find a cycle of length $3$
which separates $p$ and $q$.
We denote it by $\ell_i$.

Let $\mathcal{C}=\{\ell_i \mid 1\leq i \leq m\}$.
Since every cycle in $\mathcal{C}$ has length $3$,
any two cycles in $\mathcal{C}$ do not cross transversely on the plane,
that is, the $2$-cell regions bounded by them are disjoint or one contains the other.
(These cycles may share vertices and an edge.) 
Thus, there exists a cycle in $\mathcal{C}$ such that no other cycles in $\mathcal{C}$ embedded inside it.

Suppose that there exist two separating cycles, say $\ell_1$ and $\ell_m$,
in $\mathcal{C}$ such that no other cycles in $\mathcal{C}$ are embedded inside them.
If there are at least two separating $3$-cycles containing $x_1$ in $H_1$,
we take $\ell_1$ to be the innermost one.
Similarly, we take $\ell_m$ to be the innermost separating $3$-cycles containing $x_m$ in $H_m$.
Let $y_1$ and $y_m$ be vertices inside $\ell_1$ and $\ell_m$, respectively.
If $y_1$ belongs to $V_2$, we let $H = G[V_1\cup V_3 \cup \{y_1\}]$.
Since there is no vertex in $V_3$ inside $\ell_1$ and $\ell_1$ is the innermost separating cycle containing $x_1$,
there is no separating cycle containing $y_1$ in $H$.
Thus, $H$ is an outerplane graph. 
It follows from the inequality $\vert V(H)\vert = \vert V_1 \vert + \vert V_3 \vert + 1 \geq \lceil\frac{2n+2}{3}\rceil$ that it is a desired subgraph.
Similarly, if $y_m$ belongs to $V_2$, we can find a desired subgraph.
Thus, $y_1$ and $y_m$ belong to $V_1$.
Moreover, we may assume that there is no vertex in $V_2$ both inside $\ell_1$ and inside $\ell_m$.

Since $G$ is a $2$-tree,
every vertex of $G$ is contained in some $3$-cycle in $G$.
Since every vertex inside $\ell_1$ belongs to $V_1$,
they are independent, and hence
they must be adjacent to the two vertices in $\ell_1\cap(V_2\cup V_3)$.
However, $\ell_1$ is the innermost $3$-cycle.
It implies that $y_1$ is the only vertex inside $\ell_1$.
Similarly, $y_m$ is the only vertex inside $\ell_m$.
Thus, the induced subgraph $H'=G[V_2 \cup V_3 \cup \{y_1,y_m\}]$ has no separating cycle, and hence it is an outerplane graph.
It follows from the inequality $\vert V(H')\vert = \vert V_2 \vert + \vert V_3 \vert + 2 \geq \lceil\frac{2n+2}{3}\rceil$ that it is a desired subgraph.

Suppose that there is only one cycle, say $\ell_1$, in $\mathcal{C}$
such that no other cycles in $\mathcal{C}$ embedded inside it.
Then, for any $2\leq i \leq m$, the $2$-cell region bounded by $\ell_i$ includes $\ell_1$,
and hence for any two cycles in $\mathcal{C}$, one is embedded inside the other.
Thus, there exists a cycle, say $\ell_m$, in $\mathcal{C}$
such that no other cycles in $\mathcal{C}$ embedded outside it.
From the same arguments as in the case where there is no cycle in $\mathcal{C}$ embedded inside $\ell_1$ or $\ell_m$,
we can obtain an induced outerplane subgraph of order at least $\lceil\frac{2n+2}{3}\rceil$:
we take $\ell_1$ (resp.~$\ell_m$) to be the innermost (resp.~outermost) separating $3$-cycles containing $x_1$ (resp.~$x_m$) in $H_1$ (resp.~$H_m$).
\qed

We now construct a family of outerplanar graphs which can be embedded on the plane
so that this embedding achieves the lower bound given by Theorem~\ref{th:OuterplaneSubgraphIn2tree}.
Let $G_3$ be the graph consisting of disjoint union of $k$ $3$-cycles $C_1,C_2,\ldots , C_k$ and three isolated vertices $v_1,v_2$ and $v_3$.
We embed $G_3$ on the plane so that
$C_i$ is embedded inside $C_{i+1}$ for $1\leq i \leq k-1$, and
$v_2$ lies outside $C_k$ and $v_1$ and $v_3$ lie inside $C_1$.
It is easy to check that any induced outerplane subgraph of $G_3$ has at most $\lceil\frac{2(3k+3)+2}{3}\rceil=2k+3$ vertices.
It can be easily verified that the outerplanar graphs $G_2=G_3-v_3$ and $G_1=G_2-v_2$ also achieve the lower bound given by Theorem~\ref{th:OuterplaneSubgraphIn2tree}. 

\begin{figure}[ht]
\center
\includegraphics[scale=0.3]{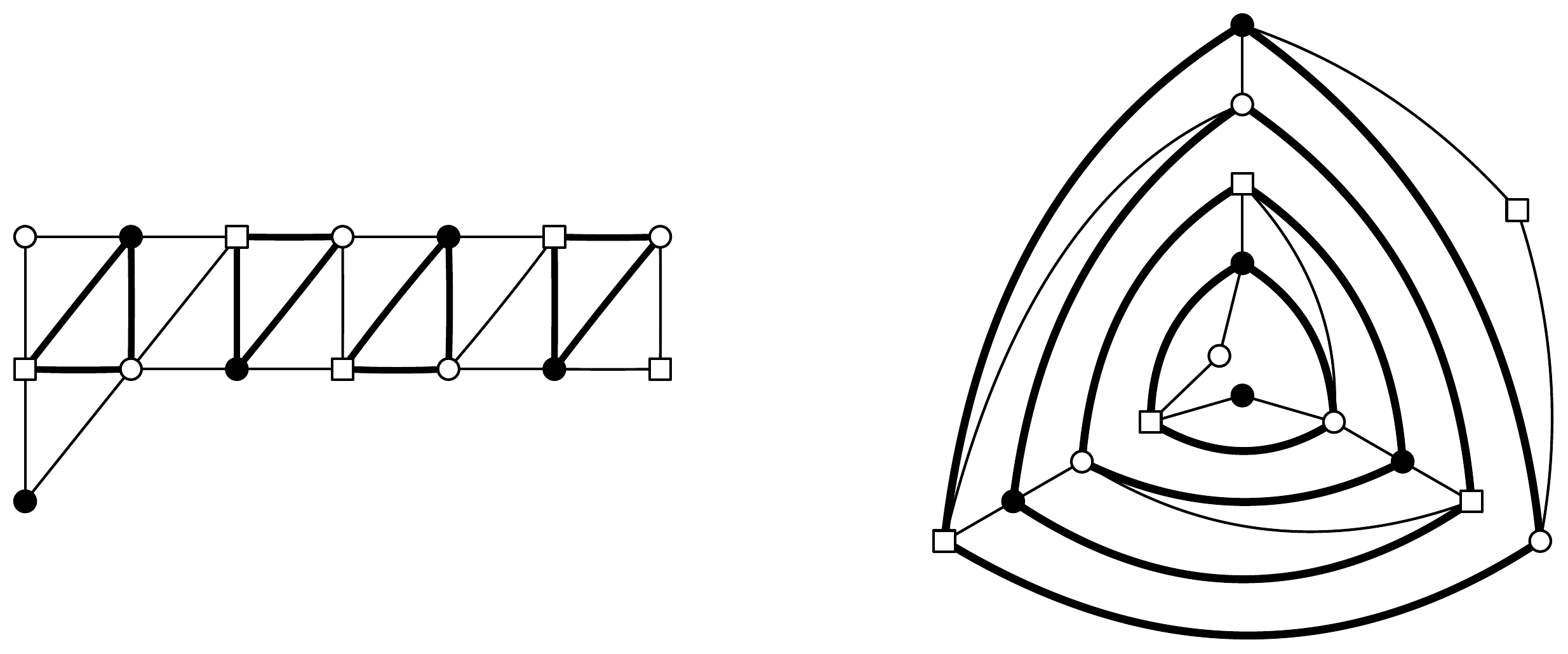}
\caption{A maximal outerplanar graph attaining the bound given by Theorem~\ref{th:OuterplaneSubgraphIn2tree}}
\label{fig:tightMaxOuterplanar}
\end{figure}

The left and right graphs in Figure~\ref{fig:tightMaxOuterplanar} represent the same outerplanar graph,
which is obtained by adding some edges to $G_3$.
The right graph shows an embedding mentioned above.
(The bold lines represent edges of $k$ $3$-cycles $C_1,C_2,\ldots , C_k$)

\section{Proof of Theorem~\ref{thm:counterexample}}\label{sec:counterexample_cd}
 
In this section, we present an infinite family of plane Eulerian triangulations that serve as counterexamples to Question~\ref{Q:connected}.

\medskip
\noindent
{\bf Construction}: 
For each integer $k \ge 1$,
we construct a plane graph $G_k$ as follows.

\begin{figure}[ht]
\center
\includegraphics[scale=0.35]{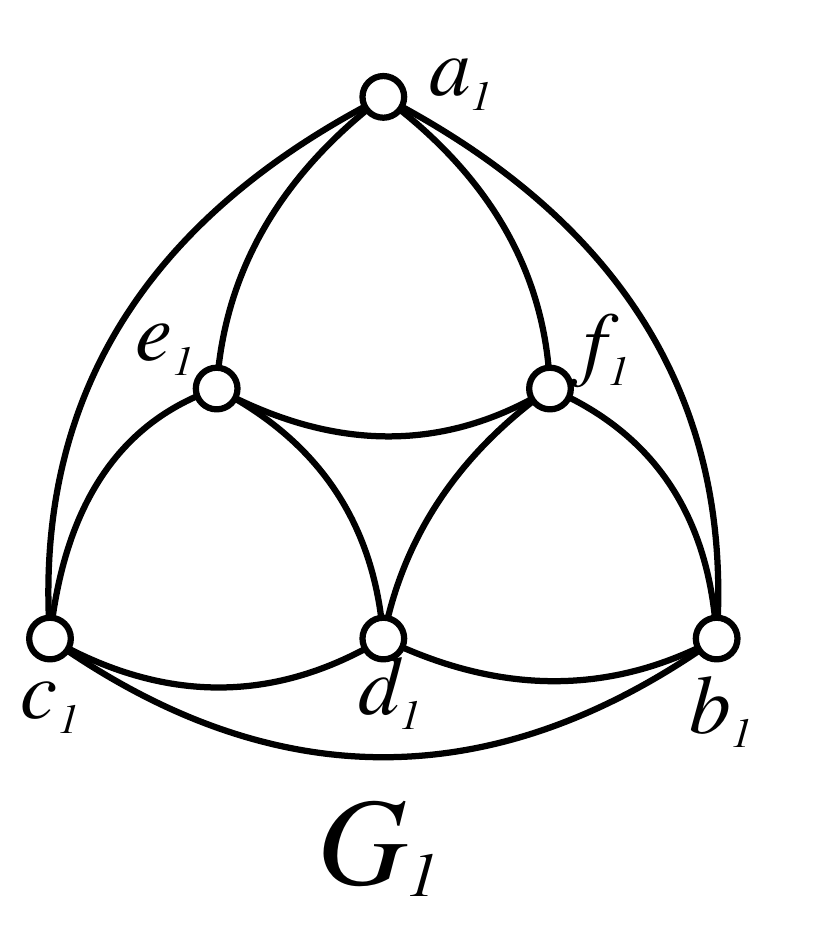}
\hspace{2em}
\includegraphics[scale=0.35]{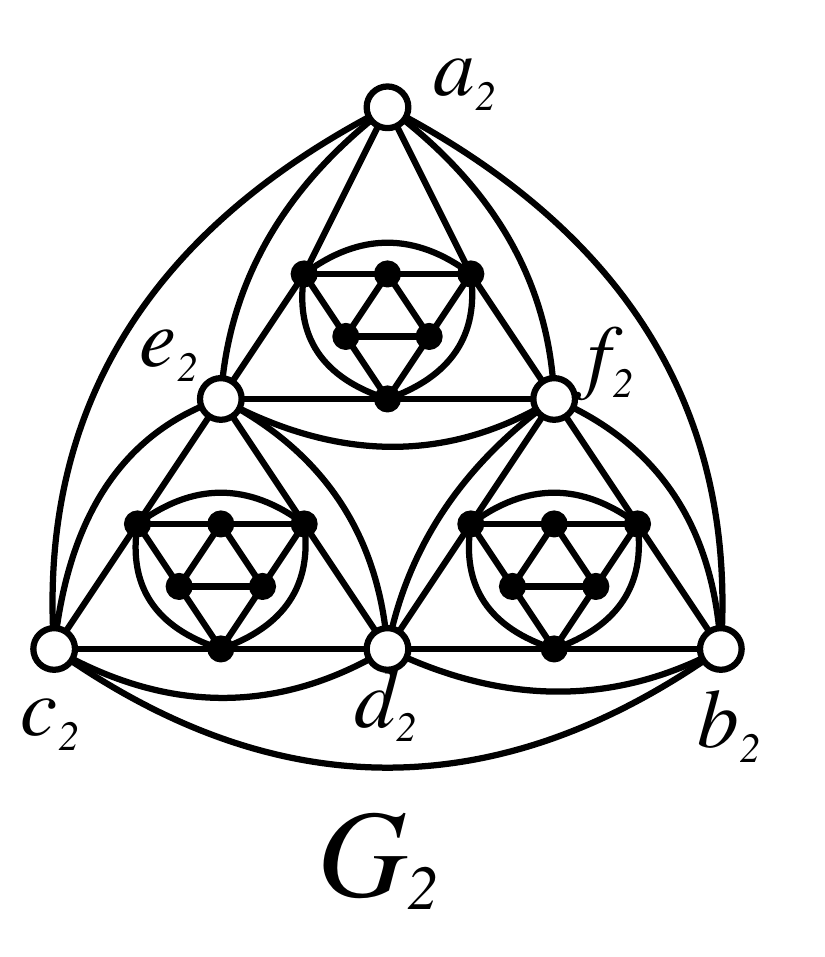}
\caption{Graphs $G_1$ and $G_2$}
\label{fig:graphG1andG2}
\end{figure}

The plane graph $G_1$, shown on the left in Figure~\ref{fig:graphG1andG2},
is the octahedron graph with six vertices $a_1,b_1,c_1,d_1,e_1$ and $f_1$,
where the outerface is bounded by a $3$-cycle $a_1b_1c_1$.
The plane graph $G_2$ is shown on the right in Figure~\ref{fig:graphG1andG2}.
To construct $G_2$, we begin with the octahedron graph with six vertices $a_2,b_2,c_2,d_2,e_2$ and $f_2$,
where the outerface is bounded by a $3$-cycle $a_2b_2c_2$.
For each of the three faces $a_2f_2e_2$, $f_2b_2d_2$ and $e_2d_2c_2$,
we insert a copy of $G_1$ into its interior.
For each of these faces, we add six edges so that
the three vertices on its boundary together with the three vertices corresponding to $a_1,b_1$ and $c_1$ in the inserted copy of $G_1$ induce an octahedron.

\begin{figure}[ht]
\center
\includegraphics[scale=0.4]{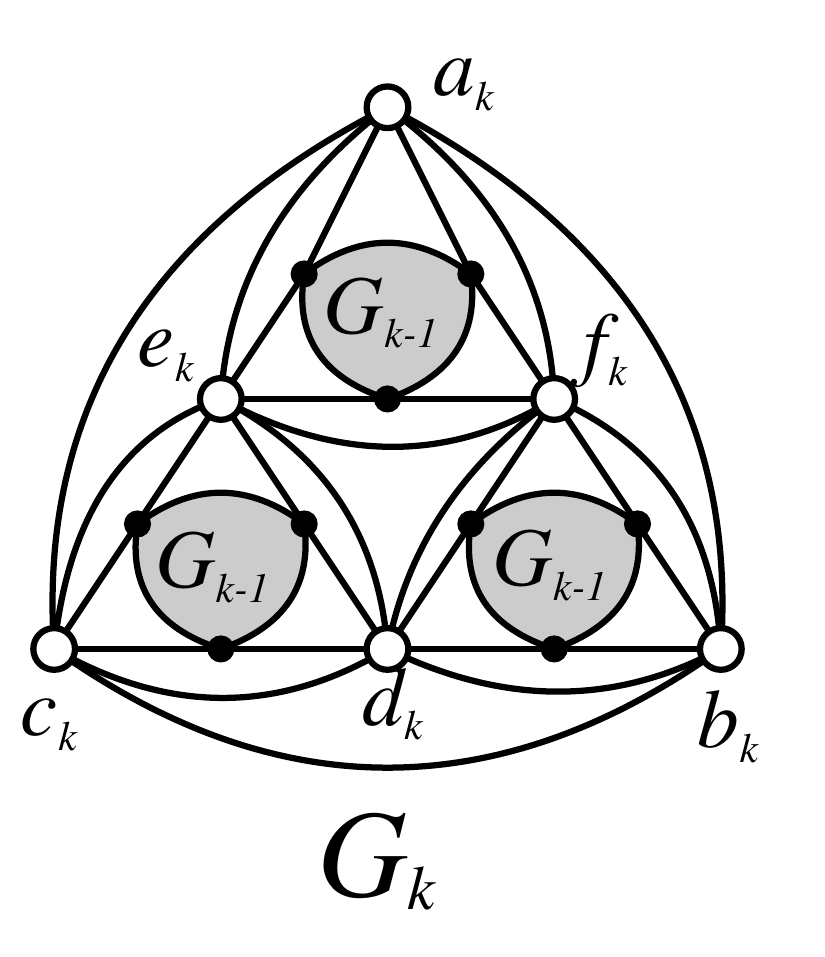}
\caption{Graph $G_k$}
\label{fig:graphGk}
\end{figure}

Following the construction of $G_2$ from $G_1$,
the graph $G_k$ can be recursively defined from $G_{k-1}$ for any $k\geq2$. 
To construct $G_k$, 
we begin with the octahedron graph with six vertices $a_k,b_k,c_k,d_k,e_k$ and $f_k$,
where the outerface is bounded by a $3$-cycle $a_kb_kc_k$.
For each of the three faces $a_kf_ke_k$, $f_kb_kd_k$ and $e_kd_kc_k$,
we insert a copy of $G_{k-1}$ into its interior.
For each of these faces, we add six edges so that
the three vertices on its boundary together with the three vertices corresponding to $a_{k-1},b_{k-1}$ and $c_{k-1}$ in the inserted copy of $G_{k-1}$ induce an octahedron.
See Figure~\ref{fig:graphGk} for an illustration of $G_k$.

\begin{figure}[ht]
\center
\includegraphics[scale=0.35]{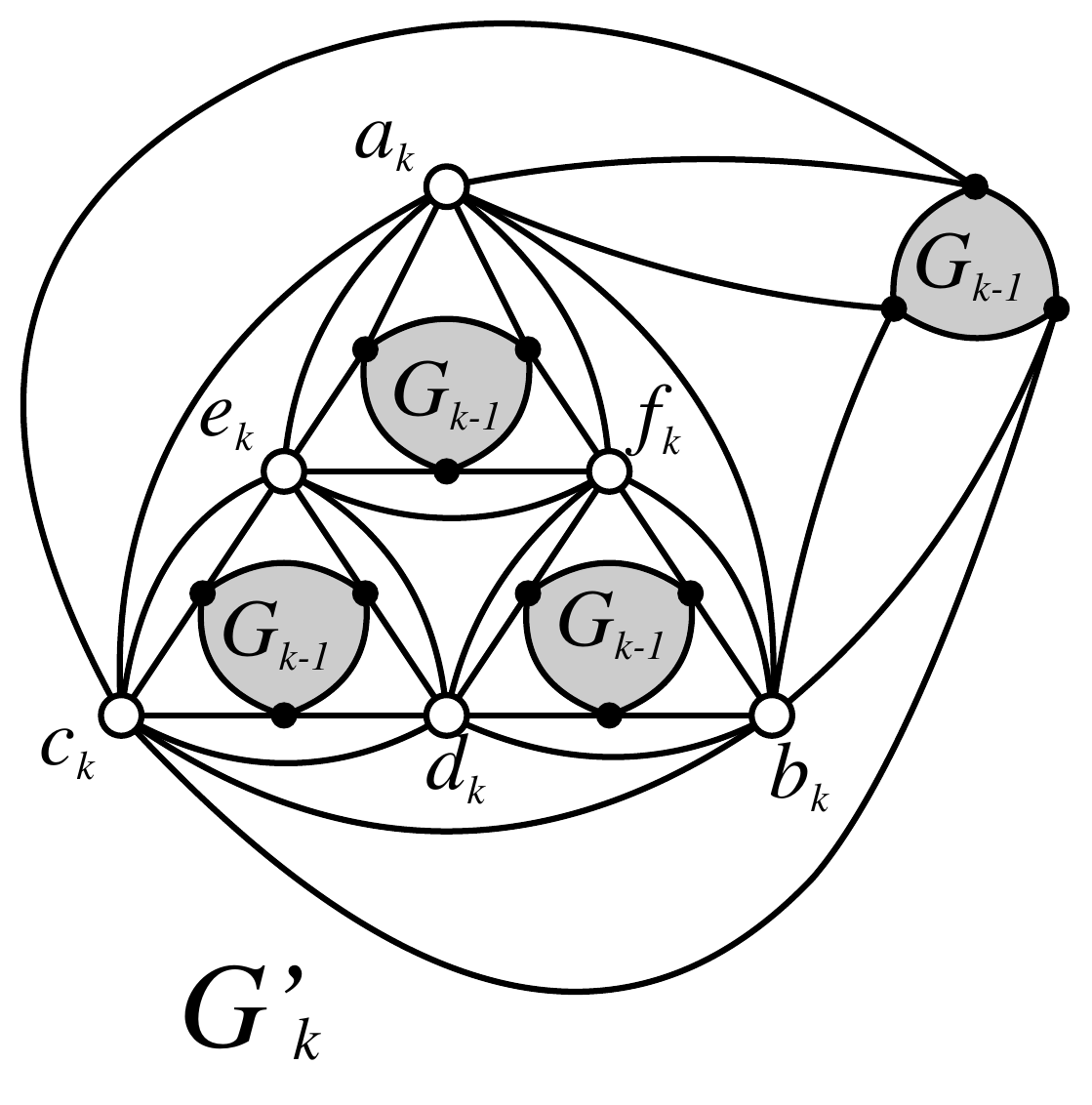}
\caption{Graph $G'_k$}
\label{fig:graphGkk}
\end{figure}

Furthermore, for $k\geq 2$, we construct the graph $G'_k$ from the octahedron graph with six vertices $a_k,b_k,c_k,d_k,e_k$ and $f_k$
by inserting a copy of $G_{k-1}$ into each of the ``four" faces $a_kf_ke_k$, $f_kb_kd_k$, $e_kd_kc_k$, and $a_kb_kc_k$
(see Figure~\ref{fig:graphGkk}).

\begin{obs}\label{obs:numOfVertOfVG_k}
For any $k\geq1$, $|V(G_k)|=3^{k+1}-3$ and for any $k'\geq2$, $|V(G'_{k'})| = 4\cdot3^{k'}-6$.
\end{obs}

\proof{}
It is easy to check that $|V(G_k)|=3|V(G_{k-1})|+6$.
Since $|V(G_1)|=6$, a simple induction yields $|V(G_k)|=3^{k+1}-3$ for each $k\geq 1$.
Moreover, we have $|V(G'_k)| = 4|V(G_{k-1})|+6$ for each $k\geq 2$,
and hence $|V(G'_k)| = 4\cdot3^k-6$.
\qed

The graph $G_k$ (or $G'_k$) contains many octahedra as subgraphs.
For an octahedron $O$, which is a subgraph of $G_k$ (or $G'_k$),
we say that $O$ is \emph{separating} in $G_k$
if the graph $G_k-V(O)$ (or $G'_k-V(O)$) is not connected.

\begin{obs}\label{obs:decomposition}
    For any $k\ge 2$, the graph $G'_k$ can be decomposed into $2\cdot3^{k-1}-1$ octahedra.
    In particular, $4\cdot3^{k-2}$ of these octahedra are non-separating. 
\end{obs}

\proof{}
One can also construct $G_k$ from $G_{k-1}$ by inserting an octahedron into each of the $3^{k-1}$ pairwise disjoint faces of $G_{k-1}$.
Then, $G_k$ can be decomposed into $\frac{3^k - 1}{2} (=1+3+\cdots+3^{k-1})$ dosjoint octahedra.
In addition, all the octahedra inherited from $G_{k-1}$ are separating in $G_k$,
while all the $3^{k-1}$ newly inserted octahedra are non-separating in $G_k$.
Therefore, it is easy to check that $G'_k$ can be decomposed into $2\cdot3^{k-1}-1$ dosjoint octahedra,
and $4\cdot3^{k-2}$ of these octahedra are non-separating in $G'_k$.
\qed

\begin{figure}[ht]
\center
\includegraphics[scale=0.3]{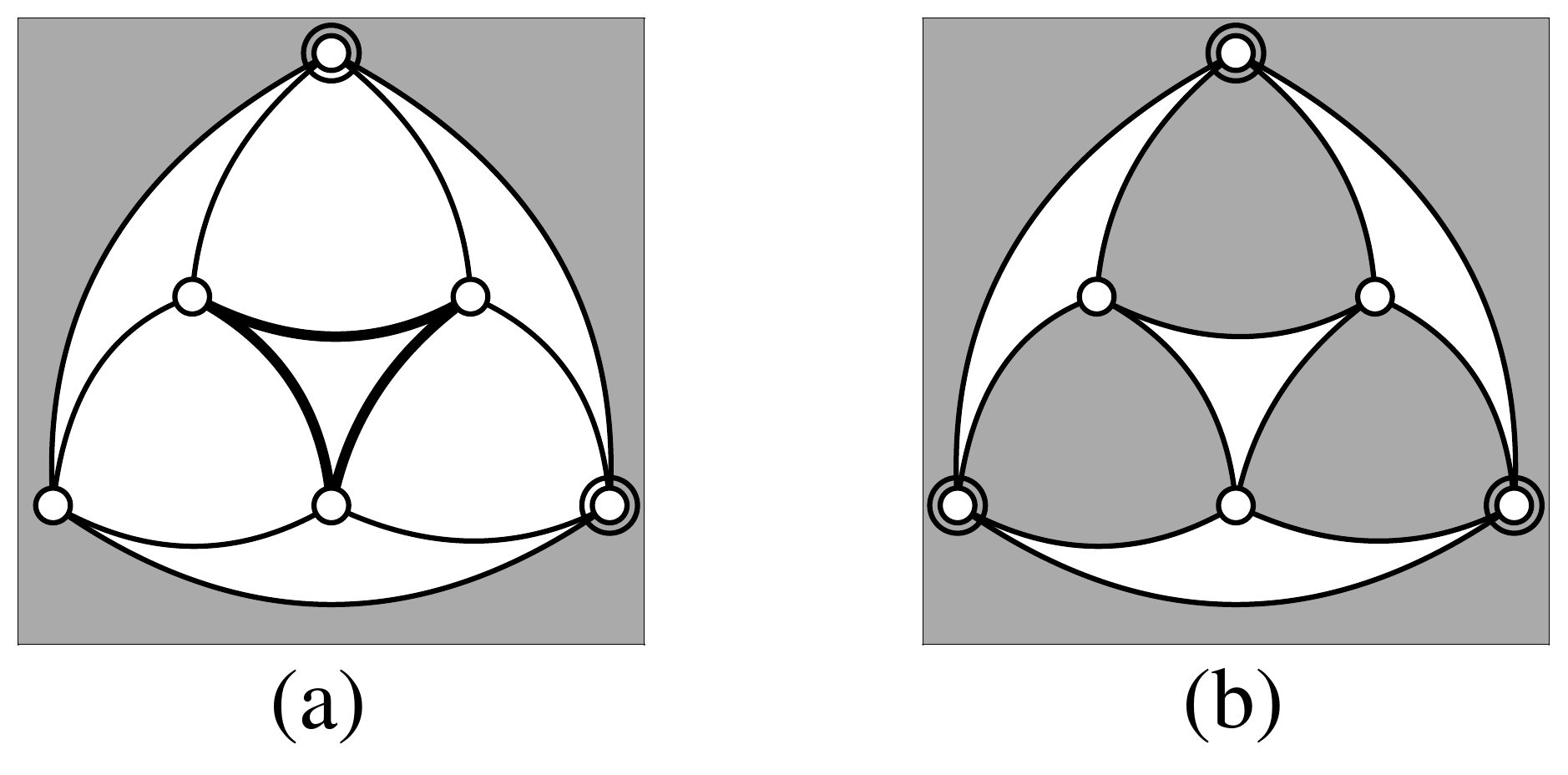}
\caption{(a) Non-separating octahedron subgraph $O$ and (b) separating octahedron subgraph $O'$.
The vertices enclosed in circles belong to $S$.}
\label{fig:non-sepAndSep}
\end{figure}

We now evaluate the connected domination number of $G'_k$.
Let $S$ be a connected dominating set of $G'_k$.
By Observation~\ref{obs:decomposition},
$G'_k$ can be decomposed into $2\cdot3^{k-2}-1$ separating octahedra and $4\cdot3^{k-2}$ non-separating ones.
For each such non-separating octahedron $O$,
the plane graph $G'_k-V(O)$ consists of a single connected component,
which is embedded in one of the faces of $O$;
this face is shown as the gray region on the left side of Figure~\ref{fig:non-sepAndSep}.
This implies that $O$ contains a facial 3-cycle $C$ that is not connected by any edge to the other octahedra in the decomposition,
where this 3-cycle $C$ is represented by the bold triangle on the left side of Figure~\ref{fig:non-sepAndSep}.
To dominate the vertices on $C$ and keep $G[S]$ connected,
$O$ must contain at least two vertices from $S$.

On the other hand, in the case of separating octahedron $O'$,
the plane graph $G'_k-V(O')$ consists of four connected components,
and each component is embedded in a distinct face of $O'$,
as illustrated on the right side of Figure~\ref{fig:non-sepAndSep}.
Since each component of $G'_k-V(O')$ contains at least one non-separating octahedron,
it contains at least two vertices of $S$.
This implies that $O'$ must contain at least three vertices of $S$
to ensure that $G[S]$ is connected.

Therefore, we see that
$$\gamma_c(G'_k)\geq  2(4\cdot3^{k-2}) + 3(2\cdot3^{k-2}-1) = 14\cdot3^{k-2} -3= \frac{7}{18}|V(G'_k)|-\frac{2}{3}.$$

Furthermore, it can immediately be checked that $G'_k$ is Eulerian.
This completes the proof of Theorem~\ref{thm:counterexample}.

\begin{rem}
\label{rem:final}
\rm
We further observe that the constructed graph $G'_k$ attains the bound in Theorem~\ref{thm:counterexample},
that is, $\gamma_c(G'_k) = \frac{7}{18}|V(G'_k)| - \frac{2}{3}$.
By Observation~\ref{obs:decomposition},
$G'_k$ can be decomposed into $4\cdot3^{k-2}$ separating octahedra and $2\cdot3^{k-2}-1$ non-separating ones.
For each octahedron, if a dominating set $S$ is chosen as in Figure~\ref{fig:non-sepAndSep}, then it is easy to see that $G[S]$ is connected.
\end{rem}

\begin{rem}
\label{rem:final2}
\rm
We also see that $s_{o'}(G'_k) = \frac{11}{18}|V(G'_k)| + \frac{2}{3}$. 
By Remark~\ref{rem:final},
we have $s_{o'}(G'_k) \ge \frac{11}{18}|V(G'_k)| + \frac{2}{3}$,
and hence it suffice to show that $s_{o'}(G'_k) \leq \frac{11}{18}|V(G'_k)| + \frac{2}{3}$.
Let $H$ be a largest induced outerplane subgraph of $G'_k$.
By Observation~\ref{obs:decomposition},
$G'_k$ can be decomposed into $2\cdot3^{k-2}-1$ separating octahedra and $4\cdot3^{k-2}$ non-separating ones.
For each such non-separating octahedron $O$,
$H$ contains at most four vertices of $O$.
Otherwise, five or more such vertices would induce a $K_4$-minor.
Next, consider the case of a separating octahedron $O'$.
If $H$ contains four or more vertices of $O'$, then it must contain a cycle $C$ such that both the interior and the exterior of $C$ contain vertices of $H$.
This contradicts the assumption that $H$ is an outerplane subgraph.
Thus, $H$ can contain at most three vertices of $O'$.
Therefore, we have
$$ s_{o'}(G'_k) \leq 4(4\cdot3^{k-2}) + 3(2\cdot3^{k-2}-1) 
= \frac{11}{18}|V(G'_k)| + \frac{2}{3}. $$
\end{rem}

\section*{Declaration of competing interest}

The authors declare that they have no known competing financial interests or personal relationships that could have appeared to influence the work reported in this paper.

\section*{Data availability}

No data was used for the research described in the article.

\section*{Acknowledgments}

This work was supported by the Research Institute for Mathematical Sciences, an International Joint Usage/Research Center located in Kyoto University.

The authors are grateful to Proffesor Kenta Noguchi for helpful comments that improved the manuscript.



\begin{thebibliography}{99}

\bibitem{AW87}
J.~Akiyama and M.~Watanabe,
Maximum induced forests of planar graphs,
\emph{Graphs Combin.} {\bf 3} (1987) 201--202.

\bibitem{AB79}
M.~O.~Albertson and D.~M.~Berman,
A conjecture on planar graphs,
In: J.A.~Bondy, U.S.R.~Murty (eds.),
\emph{Graph Theory and Related Topics},
Academic Press (1979) p.357.

\bibitem{AEFG16}
P.~Angelini, W.~Evans, F.~Frati and J.~Gudmundsson,
SEFE without mapping via large induced outerplane graphs in plane graphs,
\emph{J. Graph Theory} {\bf 82} (2016) 45--64.

\bibitem{BLS17}
G.~Borradaile, H.~Le and M.~Sherman-Bennett,
Large induced acyclic and outerplanar subgraphs of 2-outerplanar graph,
\emph{Graphs Combin.} {\bf 33} (2017) 1621--1634.

\bibitem{B79}
O.~V.~Borodin,
On acyclic colorings of planar graphs,
\emph{Discrete Math.} {\bf 25(3)} (1979) 211--236.

\bibitem{BDHMO23+}
P.~Bose, V.~Dujmovi\'{c}, H.~Houdrouge, P.~Morin and S.~Odak,
Connected dominating sets in triangulations,
arXiv: https://arxiv.org/abs/2312.03399

\bibitem{BMNS22}
P.~Bradshaw, T.~Masa\v{r}\'{i}k, J.~Novotn\'{a}, and L.~Stacho,
Robust connectivity of graphs on surfaces,
\emph{SIAM Journal on Discrete Mathematics}, {\bf 36(2)} (2022) 1416--1435.

\bibitem{BP25}
F.~Bryant and E.~Pavelescu,
Connected domination in plane triangulations,
\emph{Involve, a Journal of Math.} {\bf 18(3)} (2025) 555--566.

\bibitem{CP13}
G.~G.~Chappell and M.~J.~Pelsmajer,
Maximum induced forests in graphs of bounded treewidth,
\emph{Electron. J. Combin.} {\bf 20(4)} (2013) P8.

\bibitem{CRR2024}
A.B.G.~Christiansen, E.~Rotenberg and D.~Rutschmann,
Triangulations admit dominating sets of size $2n/7$,
\emph{Proceedings of the 2024 Annual ACM-SIAM Symposium on Discrete Algorithms (SODA)}
(2024) 1194--1240.

\bibitem{CGHHMMT2021}
M.~Claverol, A.~Garc\'{i}a, G.~Hern\'{a}ndez, C.~Hernando, M.~Maureso, M.~Mora and J.~Tejel,
Total domination in plane triangulations,
\emph{Discrete Math.} {\bf 344} (2021) \#112179.

\bibitem{DF25}
M.~D'Elia and F.~Frati,
Large induced subgraphs of bounded degree in outerplanar and planar graphs,
arXiv: https://arxiv.org/pdf/2412.14784

\bibitem{DMP19}
F.~Dross, M.~Montassier and A.~Pinlou,
A lower bound on the order of the largest induced linear forest in triangle-free planar graphs
\emph{Discrete Math.} {\bf 342} (2019) 943--950.

\bibitem{DW12}
D.~Z.~Du and P.~J.~Wan,
Connected dominating set: theory and applications (Vol. 77),
Springer Science and Business Media (2012).




\bibitem{FGR02}
G.~Fertin, E.~Godard and A.~Raspaud,
Minimum feedback vertex set and acyclic coloring,
\emph{Information Processing Letters},
{\bf 84} (2002) 131--139.

\bibitem{HHH23}
T.~W.~Haynes, S.~T.~Hedetniemi and M.~A.~Henning (eds.),
Domination in graphs: Core concepts. Cham: Springer (2023).

\bibitem{HL91}
S.~T.~Hedetniemi and R.~C.~Laskar,
Topics on domination. Elsevier (1991).

\bibitem{H22}
M.~A.~Henning,
Bounds on domination parameters in graphs: a brief survey,
\emph{Discuss. Math. Graph Theory} {\bf 42} (2022) 665--708.

\bibitem{HY2013}
M.A.~Henning and A.~Yeo,
Total domination in graphs,
New York: Springer, 2013.

\bibitem{H90}
K.~Hosono,
Induced forests in trees and outerplanar graphs,
\emph{Proc. Fac. Sci. Tokai Univ.} {\bf 25} (1990) 27--29.

\bibitem{KP10}
E.~King and M.~Pelsmajer,
Dominating sets in plane triangulations,
\emph{Discrete Math.} {\bf 310} (2010) 2221--2230.

\bibitem{Le18}
H.~Le,
A better bound on the largest induced forests in triangle-free planar graph,
\emph{Graphs Combin.} {\bf 34} (2018) 1217--1246.

\bibitem{MT96}
L.~R.~Matheson and R.~E.~Tarjan,
Dominating sets in planar graphs,
\emph{European J. Combin.} {\bf 17} (1996) 565--568.

\bibitem{MY2025+}
N.~Matsumoto and T.~Yashima,
Large induced subgraph with a given pathwidth in outerplanar graphs,
arXiv: https://arxiv.org/abs/2505.23162

\bibitem{NZ2024}
K.~Noguchi and C.T.~Zamfirescu,
Spanning trees for many different numbers of leaves,
\emph{Discrete Math. Theor. Comput. Sci.} {\bf 26:3} (2024) \#17.

\bibitem{Pels04}
M.~Pelsmajer,
Maximum induced linear forests in outerplanar graphs,
\emph{Graphs Combin.} {\bf 20} (2004) 121--129.

\bibitem{SW79}
E.~Sampathkumar and H.~B.~Walikar,
The connected domination number of a graph,
\emph{J. Math. Phys. Sci.} {\bf 13} (1979) 607--613.

\bibitem{S20}
S.~\v{S}pacapan,
The domination number of plane triangulations,
\emph{J. Combin. Theory Ser. B} {\bf 143} (2020) 42--64.

\bibitem{Z20}
W.~Zhuang,
Connected domination in maximal outerplanar graphs,
\emph{Discrete Appl. Math.} {\bf 283} (2020) 533--541.

\end{thebibliography}
\end{document}